\documentclass[preprint,12pt]{aastex}

\usepackage{epstopdf}

\usepackage{graphicx}

\usepackage{subfigure}

\begin{document}

%-------------------------------------%

%     ENCAPSULATED POSTSCRIPT         %

%-------------------------------------%

\def\putplot#1#2#3#4#5#6#7{\begin{centering} \leavevmode

\vbox to#2{\rule{0pt}{#2}}

\includegraphics{#1}

% e.g.,

% \putplot{psfile}{vspace}{angle}{hscale}{vscale}{hoffset}{voffset}

% with vspace in any TeX units, angle in degrees, scale in percent,

% and offset in PostScript points (72/in)

\end{centering}}

%-------------------------------------%

%    END ENCAPSULATED POSTSCRIPT      %

%-------------------------------------%

\def\Msun{M_\odot}

\def\Lsun{L_\odot}

\def\Rsun{R_\odot}

\slugcomment{Submitted to AJ}

\shorttitle{M101 Wolf-Rayet Stars}

\shortauthors{Shara et al}

\title{The Vast Population of Wolf-Rayet and Red Supergiant Stars in M101: I. Motivation and First Results}

\author{Michael~M.~Shara\altaffilmark{1}, Joanne L. Bibby\altaffilmark{1,2}, David~Zurek\altaffilmark{1}, Paul A. Crowther\altaffilmark{3}, Anthony F.J. Moffat\altaffilmark{4}, and Laurent Drissen\altaffilmark{5}}





\altaffiltext{1}{Department of Astrophysics, American Museum of Natural History, Central Park West and 79th Street, New York, NY 10024-5192}

\altaffiltext{2}{Jeremiah Horrocks Institute for Maths, Physics \& Astronomy, University of Central Lancashire, Preston, PR1 2HE, United Kingdom}

\altaffiltext{3}{Department of Physics and Astronomy, University of Sheffield, Hounsfield Road, Sheffield S3 7RH, United Kingdom}

\altaffiltext{4}{D\'epartement de Physique, Universit\'e de Montr\'eal, CP 6128 Succ. C-V, Montr\'eal, QC, H3C 3J7, Canada }

\altaffiltext{5}{D\'epartement de Physique, Universit\'e Laval, Pavillon Vachon, Quebec City, QC, G1K 7P4 Canada}

\begin{abstract}

M101 is an ideal target in which to test predictions of massive star birth and evolution. The large abundance gradient across M101 (a factor of 20) suggests that many more WR stars must be found in the inner parts of this galaxy than in the outer regions. Many H\,{\sc ii} regions and massive star-forming complexes have been identified in M101; they should be rich in WR stars, and surrounded by RSG stars. Finally, the Wolf-Rayet stars in M101 may be abundant enough for one to explode as a Type Ib or Ic supernova and/or GRB within a generation. The clear identification of the progenitor of a Type Ib or Ic supernova as a WR star would be a major confirmation of current stellar evolution theory.

Motivated by these considerations, we have used the Hubble Space Telescope to carry out a deep, He\,{\sc ii} optical narrowband imaging survey of the massive star populations in the ScI spiral galaxy M101. Combined with archival broadband images, we were able to image almost the entire galaxy with the unprecedented depth and resolution that only HST affords.

We describe the extent of the survey and our images, as well as our data reduction procedures. A detailed study of a field east of the center of M101, containing the giant star-forming region NGC 5462, demonstrates how we find candidates, their properties and spatial distribution, and how we rule out most contaminants. The spatial distributions of the WR and RSG stars near a giant star-forming complex are strikingly different. WR stars dominate the complex core, while RSG dominate the complex halo. Future papers in this series will describe and catalog more than a thousand WR and RSG candidates that are detectable in our images, as well as spectra of many of those candidates.

\end{abstract}

\keywords{galaxies: individual (M101) --- galaxies: stellar content --- stars: Wolf-Rayet ---stars: Supergiants}

\section{Introduction and Motivation}

\subsection{Starburst Regions}

The study of individual luminous stars and stellar populations in nearby giant H\,{\sc ii} regions is a prerequisite to understanding the starburst phenomenon and interpreting the observations of distant starburst galaxies and those containing starburst regions, for which only integral properties can be observed \citep{lei97}. In this context, the Hubble Space Telescope (HST) has been crucial in providing us with high-resolution images of nearby, but very dense and massive, stellar clusters which are ionizing giant H\,{\sc ii} regions (\citet{hun95}, \citet{hun96} and \citet{mal96}).

Our HST-based investigation of the stellar populations of the most luminous star-forming complexes in the nearby late-type (ScIII) spiral galaxy NGC 2403 was very fruitful \citep{dri99}, and underpinned our request for HST time to observe M101. A member of the M81 group, at a distance of 3.2 $\pm$ 0.4 Mpc \citep{fre88}, NGC 2403 is very rich in H\,{\sc ii} regions \citep{siv90}. Its abundance level and O/H radial gradient have been well established by \citet{mar96}; they are similar to those of M33 \citep{hen95}. In contrast with M33, which contains relatively modest giant H II regions, four of NGC 2403's H\,{\sc ii} regions are exceptionally bright, with H$\alpha$ luminosities L($H\alpha$) $\sim 0.8-1.5 \times 10^{40} $ erg/s, comparable to the most massive starburst region in the Local Group, the 30 Doradus complex. We also found direct evidence for the presence of Wolf-Rayet (WR) stars in five of the six giant H\,{\sc ii} regions investigated; 25 - 40 WR stars are present in the NGC 2403-I giant H\,{\sc ii} region alone. HST has also provided optical and UV spectra of individual massive stars in NGC 1569 \citep{mao01}, NGC 5398 \citep{sid06} and NGC 925 \citep{ada11}. Ground-based imagery and spectroscopy has revealed rich WR populations in M83 \citep{cro04} and NGC 5253 \citep{cro99}.

M101 (also known as NGC 5457 and the Pinwheel Galaxy) is the logical galaxy in which to extend this work. As the nearest giant grand design ScI spiral galaxy, it is brimming with H\,{\sc ii} regions, massive stars and at least 3,000 luminous star clusters \citep{bar06}. About 6500 WR stars are estimated to exist in the Milky Way \citep{sha99}; simplistically scaling up to the size and luminosity of M101 suggests a population of 10-20,000 WR stars and even more Red Supergiants (RSGs). The star formation rate (SFR) in M101 \citep{Lee2009} is probably a few times that of the Milky Way, and M101 is 50\% or more larger than our Galaxy. This, again, suggests a very substantial population of M101 WR stars. In addition, a rich treasury of (mostly) continuum imagery with HST was already in hand: over 135 ksec of HST exposures. As described below, we used this database extensively to perform the image subtractions needed to isolate the strong emission-line WR stars and very red Red Supergiants from the other stellar populations.

\subsection{Wolf-Rayet and Red Supergiant Stars}

Wolf-Rayet \citep{cro07} and Red Supergiant stars (\citet{lev10}; \citet{mey11}) are the massive stars that are easiest to identify in imaging surveys of galaxies because of their strong emission lines and extreme colors, respectively. They provide important constraints on the age of a starburst \citep{gaz12} and on the mode of star formation \citep{cro13}. Single stars with initial masses (M$_{i}>$20M$_{\odot}$) are predicted to advance to the WR phase at approximately solar metallicity. WR stars possess strong stellar winds which produce a unique, emission--line spectrum displaying broad He\,{\sc ii}\,$\lambda$4686 for hot nitrogen--rich (WN) and carbon--rich (WC) stars or C\,{\sc iii}\,$\lambda$4650+C\,{\sc iv}\,$\lambda$5808 for (initially more massive) carbon--rich (WC) subtypes. WNh stars are a unique subclass of WR stars as these are luminous WN stars that are still burning hydrogen on the main sequence \citep{dek98}, hence they are fundamentally different from their helium burning \textit{classical} cousins.

Since the hydrogen--rich envelope has been removed from classical WN stars it follows that they are probably the progenitors of at least a subset of H--poor Type Ib SN. Similarly, the removal of both the hydrogen and helium envelopes from WC stars should correspond to the absence of both these elements in the spectra of Type Ic SN.  However, the WR-Type Ibc SN question remains unresolved as, to date, \textit{no direct detection of a Type Ib or Ic SN progenitor has been obtained.} \citep{eld13} have recently claimed that 12 SNIbc progenitors are invisible to as faint as absolute B, V and R magnitudes of -4 to -5.

In contrast, RSG are predicted to arise from the evolution of less massive (8 -- 20 M$_{\odot}$) stars and are therefore expected to appear later in the life of a starburst. Evolutionary models predict that single massive stars with M$_{i}\sim$8--20M$_{\odot}$ end their lives during the Red Supergiant (RSG) phase as H--rich Type II core--collapse supernovae (ccSNe). HST broad--band pre-SN imaging has been able to confirm the RSG--Type II SN connection, particularly for SN 2003gd \citep{sma09}. However, the highest mass RSG progenitor to date is only $\sim$16M$_{\odot}$ (Smartt 2009), making these limits uncertain.

Models indicate that in instantaneous starbursts of low metallicity, these two populations are well separated in time, since only the most extreme stars (M$_{i}$ $\geq$ 50 {M$_{\odot}$) can shed enough mass to reach the WR stage. In regions of high metallicity, however, the simultaneous presence of WR and RSG stars can be expected for a short period of time, since lower mass stars ( $ \sim 25$ {M$_{\odot}$) can also become WR after having spent some time as RSG \citep{mae94}. WR and RSG are observed to coexist in the massive Galactic cluster Westerlund 1 \citep{cla05}. One of our key goals is to directly test this prediction in a single galaxy: M101. \citet{ken03} have shown that in M101, over the galactocentric range 6-41 kpc, oxygen abundances are well fitted by an exponential distribution from approximately 1.3 (O/H) solar in the center to 1/15 (O/H) solar in the outermost regions. Equivalently, log O/H +12 = 8.8 in the center of M101, and 7.5 in that galaxy's outer regions. Observing across the entire range of galactocentric distances in M101 (from 0 to 50 kpc) to measure how the absolute numbers and WR/RSG ratio changes across the galaxy is an equally important goal of our study.

\subsection{Star Clusters}

An early HST--based investigation by \citet{dri93} targeted the WR population of M33 using narrow--band $\lambda$4686 imaging surveys. NGC 604 is the largest giant H\,{\sc ii} region in the nearby star-forming galaxy M33 which lies at a distance of only d$\sim$0.8 Mpc \citep{sco09}. Ground--based imaging with seeing $\sim$1.2\arcsec ~revealed that NGC 604 has a moderate WR population \citep{dri91}. However this was significantly increased via high spatial resolution HST narrow--band imaging \citep{dri93}, identifying the fainter WR population which corresponds to the lowest mass WR stars. At the core of each of the two most luminous giant H II regions in NGC 2403 lies a luminous, compact object \citep{dri99}. 

The discovery that very dense, massive star clusters form at the cores of all types of starbursts led to the suggestion that globular clusters were once located at the cores of massive starbursts (Meurer 1995; Whitmore \& Schweizer 1995; Ho \& Filippenko 1996a). The cases of HD 97950 (the ionizing core of NGC 3603 \citep{dri95}), R136 \citep{mof94} and \citep{cro10}, NGC 2363 \citep{dri00}, NGC 2403-I and NGC 2403-II \citep{dri99}, M31 \citep{neu12} and M33 \citep{neu11} show that massive compact stellar clusters also form in more normal galaxies as well. How much more common are they in a very massive, actively star-forming galaxy like M101?

%The discovery that very dense, massive star clusters form at the cores of all types of starbursts led to the suggestion that globular clusters were once located at the core of massive starbursts (Meurer 1995; Whitmore \& Schweizer 1995; Ho \& Filippenko 1996a). The cases of HD 97950 (the ionizing core of NGC 3603 \citep{dri95}), R136 \citep{mof94}, NGC 2363 \citep{dri00}, NGC 2403-I and NGC 2403-II, M31, M33 and the LMC (\citet{neu11}), \citep{neu12} and \citet{cro10} show that massive compact stellar clusters also form in more normal galaxies as well. How much more common are they in a very massive, actively star-forming galaxy like M101?

A striking feature in the NGC 2403 clusters is that RSG stars are mainly present over a more extended halo, while the young blue stars and most WR stars are in or close to a compact core. Stars more massive than $\sim 25$ M$_{\odot}$ are not expected to go through a RSG phase before becoming WR stars.  For $M_{i}$ $\leq$ 20 M$_{\odot}$, stars evolve to RSG and explode as Type II supernovae without entering the WR phase. The timescale of these evolutionary paths is longer as one considers lower masses, hence the absence of RSG and presence of WR stars in the cores indicate that the population is dominated by very young and very massive stars. The presence of RSG in the halos signifies that we have an older mix of stars of various masses with M $\gtrapprox$ 8--15 M$_{\odot}$. The relative age spread and the spatial exclusion between RSG and WR stars are most obvious for the largest H\,{\sc ii} regions.  Although of different ages, the proximity there of the WR and RSG stars suggests a triggering link between the two populations, which is a key part of this research program. WR stars and RSG are observed to co-exist in the Milky Way's most massive compact cluster Westerlund 1 \citep{cla05}. The luminosity, inferred mass and compact nature of Westerlund 1 are comparable with those of Super Star Clusters - previously identified only in external galaxies (see e.g. \citet{bas12}, \citet{lar11},\citet{whi05} and \citet{whi11}).

\subsection{Supernova Environments}

Another key goal of this and related studies is obtaining narrow--band imaging of several nearby galaxies to produce a catalogue of $\sim$10$^4$ WR stars. When a Type Ibc SN and/or gamma ray burst \citep{geo12} eventually occurs in one of these galaxies, our catalog should reveal the WR progenitor, confirming one of the strongest predictions of stellar evolutionary theory.  We have obtained ground-based narrow-band imaging of several nearby star-forming galaxies (\citet{bib12}, \citet{bib10} and \citet{had07}) and confirmed a subset of the WR candidates with multi-object spectroscopy. Given the average lifetime of a WR star of $\sim$ 0.3\,Myr \citep{cro03}, we would expect one of the WR stars identified to produce a Type Ibc ccSN within the next few decades. Until then, we are able to compare the distribution of WR stars in their host galaxy with different types of SN to assess whether they represent a common population; and to check whether the predicted WC/WN ratio varies across galaxies as predicted by theory.

Different ccSN are seen to be located in different regions of their host galaxies. For example, Type II ccSN follow the distribution of the host galaxy light whereas Type Ic SN are preferentially located in the brightest regions, similar to long Gamma-Ray Bursts \citep{fru06}. Furthermore, Type Ib and Ic SN were found to have different spatial distributions relative to the distribution of the host galaxy light, strongly suggesting that they have different progenitors \citep{kel08}. If WN and WC stars are the progenitors of Type Ib and Ic SN, respectively, they should follow the same distribution as the corresponding supernovae. Indeed, \citet{lel10} applied this approach to spectroscopically confirmed WR stars in M83 \citep{had05} and found that WN and WC stars are located in different regions of their host galaxy. Moreover,  the distribution of WN and WC stars are most consistent with those of Type Ib and Ic SNe, respectively. Given that this paper only concentrates on one region in M101, we postpone the discussion of our candidates' distribution until the next paper in this series.

The plan of this paper is as follows. In Section 2 we describe the narrowband imaging technique and our HST observations. The data reductions, including image processing, photometry and detection limits are presented in Section 3. Source selection from our images is described in Section 4 and we address the issue of contamination by variable stars in Section 5. Our methodology for locating Red Supergiants is presented in Section 6; we also compare the distributions of the WR and RSG candidates and star clusters in this section. We briefly summarize our results in Section 7.

\section{Techniques and Observations}

The need for targeted surveys that uniquely pick out WR stars is highlighted by \citet{sma09} and by \citet{eld13}, who discussed Type Ibc SN for which broad--band pre-SN imaging exists. Not one progenitor has been identified from these dozen SNe. This is because short exposure times did not allow the images to go deep enough to detect the continuum of a WR star. Had proper narrow--band surveys been available we would almost certainly, by now, have been able to provide strong evidence for the WR-SNIbc connection.

A powerful technique to detect individual Wolf-Rayet stars in crowded fields, such as the ones in M101, consists of subtracting a continuum image, normalized in both PSF and intensity, from an image obtained with a narrow-band filter centered on the He\,{\sc ii} $\lambda$4686 emission line. One reason for this is that WR stars are much brighter in filters sensitive to their strong, broad emission lines, particularly He\,{\sc ii}$\lambda$4686, than their continua, by up to 3 magnitudes \citep{mjo98}. Hence, WR stars can be easily identified from specific narrow--band images but are difficult to detect in broad--band images. Moreover, WR stars detected in broad--band images alone cannot be distinguished from other blue supergiants. This narrowband-broadband technique has been successfully used on WFPC2 F469N images of giant H\,{\sc ii} regions in galaxies such as M33 \citep{dri93} and NGC~2403 \citep{dri99}, detecting both isolated, individual WR stars and WR stars in unresolved clusters that include only a very small fraction of WR stars.

M101 lies at a distance of 6.4\,Mpc \citep{sha11}, almost twice as far as NGC~2403 \citep{fre88}, so that $\sim$ 4$\times$ longer exposures are needed to reach similar magnitudes. In Cycle 17 we obtained HST/Wide Field Camera 3 (WFC3) pointings of 2 orbits per M101 field, under program ID 11635 (PI. Shara), with a total exposure time of 6106~seconds per field. This permitted us to image to a similar depth in M101 as we achieved in NGC~2403. We note that the systemic redshift of M101 (+372 km/s) shifts the center of the He\,{\sc ii} $\lambda$4686 emission line to $\lambda$4692~\AA. This has virtually no effect on the detectability of WR stars in this galaxy, since the F469N filter transmission curve is fortuitously centered at 4693 \AA, and the filter sensitivity is nearly constant from 4680 to 4710 \AA, with a FWHM of 50 \AA. It is true that the filter will capture essentially all HeII 4686 but exclude most NIII 4640, with only
the red half of CIII 4650 included.

We used 18 pointings to cover the large majority of M101. Some gaps between CCD chips are inevitable. The fields covered are shown in Figure \ref{m101_wfc3_view}. The very different orientation of one of the pointings was necessary to provide a guide star for the observations, albeit resulting in overlap with another pointing. The coverage of our WFC3 images was selected based on the availability of deep archival continuum imaging. We used F435W, F555W and F814W ACS/WFC images, taken under program ID 9490 (PI. Kuntz) and ID 10918 (PI. Freedman), to represent continuum images so that only additional narrow-band F469N images were needed.

\begin{figure}

\centering

\includegraphics[width=0.7\columnwidth]{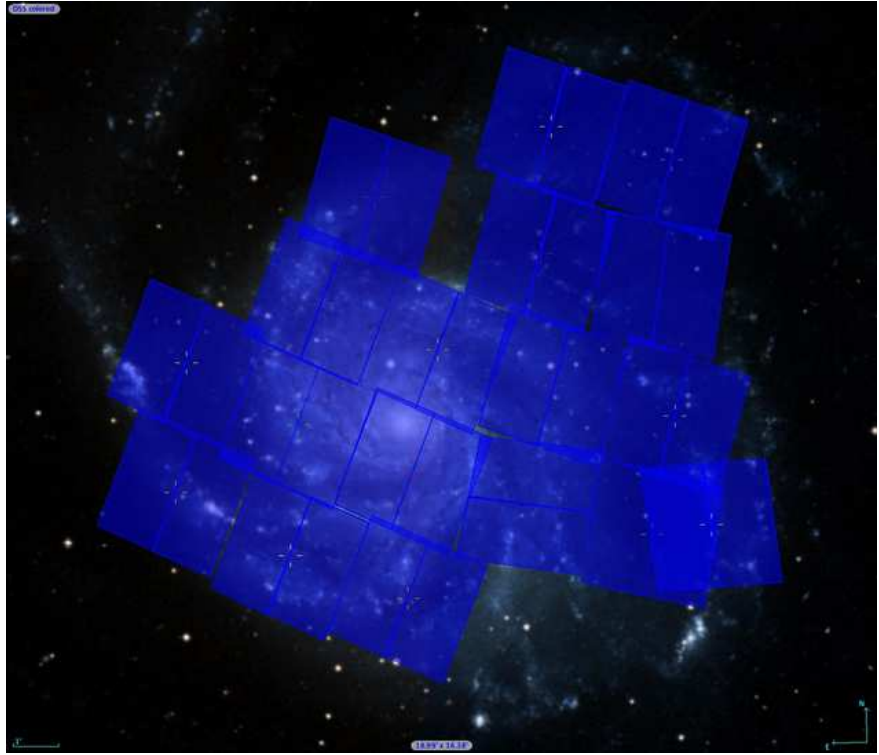}

\caption{Digitized Sky Survey image of M101 with the 18 HST/WFC3 pointings overlaid. North is up and East to the left of the image. The field shown is approximately 20 arcmin$^{2}$ and each WFC3 field highlighted is 2.7$\times$2.7 arcmin.}

\label{m101_wfc3_view}

\end{figure}

\section{Data Reductions} 

\label{reduction}

Each WFC3 pointing was treated individually to ensure the best alignment with the ACS/WFC data and to make the size of each dataset more manageable. The corresponding ACS frames were drizzled with the WFC3 F469N pointing using the \textsc{multidrizzle} task within \textsc{iraf}\footnotemark \footnotetext{IRAF is distributed by the National Optical Astronomy Observatory, which is operated by the Association of Universities for Research in Astronomy (AURA) under cooperative agreement with the National Science Foundation.} to produce a coordinate system that was consistent between each different instrument and filter. Often small shifts and rotations were required to achieve the best alignment; these were calculated using \textsc{geomap}. To achieve the best match, the WFC3 data was drizzled to a spatial scale of 0.15\arcsec, corresponding to 4.65 pc at the 6.4 Mpc distance of M101. While slightly degraded from the 0.1\arcsec optimum sampling offered by HST, this was necessary to allow us to produce the best continuum subtractions possible.

In the following we describe the methods applied to all 18 pointings, and present the results of one of those pointings (M101-I). Later papers in this series, describing the remaining 17 fields, use exactly the same methodology.

\subsection{Photometry}

Once the broad-- and narrow--band images had been aligned, photometry was performed on each filter separately using the standalone code, \textsc{daophot} \citep{ste87}. A point-spread function (PSF), based on isolated, point-like stars within the field, was built and applied to all the other stars detected.  Individual zero-points from the HST literature were applied for each filter to transform the observed magnitudes into ST magnitudes and Vega magnitudes. For this study we find the Vega magnitude system to be more useful, since (under the Vega system) the F435W, F555W, and F814W filters correspond to the Johnson B, V and I filters. This enabled us to use literature color cuts. Henceforth, any magnitudes listed are Vega magnitudes unless otherwise stated.

Typical photometric errors for the narrow-band F469N images were $\pm$0.07\,mag for bright sources (m$_{F469N}$\,=\,21mag) and $\pm$0.5\,mag for the fainter sources (m$_{F469N}$\,=\,25.5\,mag); the distribution of the photometric errors relative to the source brightness for the F469N images is shown in Figure \ref{f469_errors}. For the ACS broad-band images the errors were slightly lower for sources of similar magnitudes, with $\pm$0.03\,mag (m$_{F435W}$\,=\,21mag) and $\pm$0.10\,mag (m$_{F435W}$\,=\,25.5mag). For the faintest sources in the ACS image with m$_{F435W}$\,=\,28\,mag the photometric error is typically $\pm$0.5\,mag, shown in Figure \ref{f435_errors} for the F435W image, which is consistent with the other broad-band images.

\begin{figure} 

\centering 

\includegraphics[width=0.7\columnwidth, angle=-90]{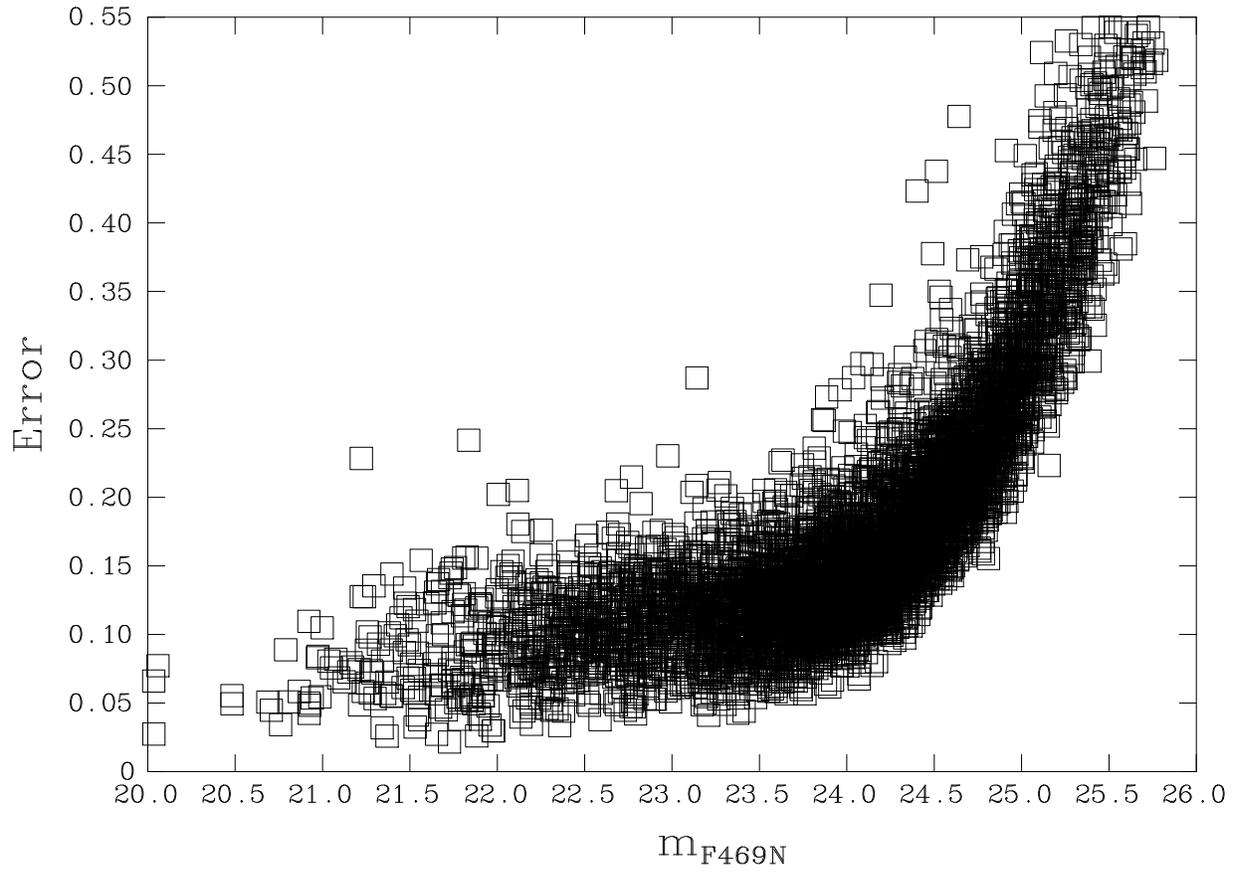} 

\caption{Photometric errors as a function of zero-point corrected, apparent (Vega) magnitude for all sources found in the WFC3/F469N image of M101-I .}

\label{f469_errors}

\end{figure}

\begin{figure} 

\centering 

\includegraphics[width=0.7\columnwidth, angle=-90]{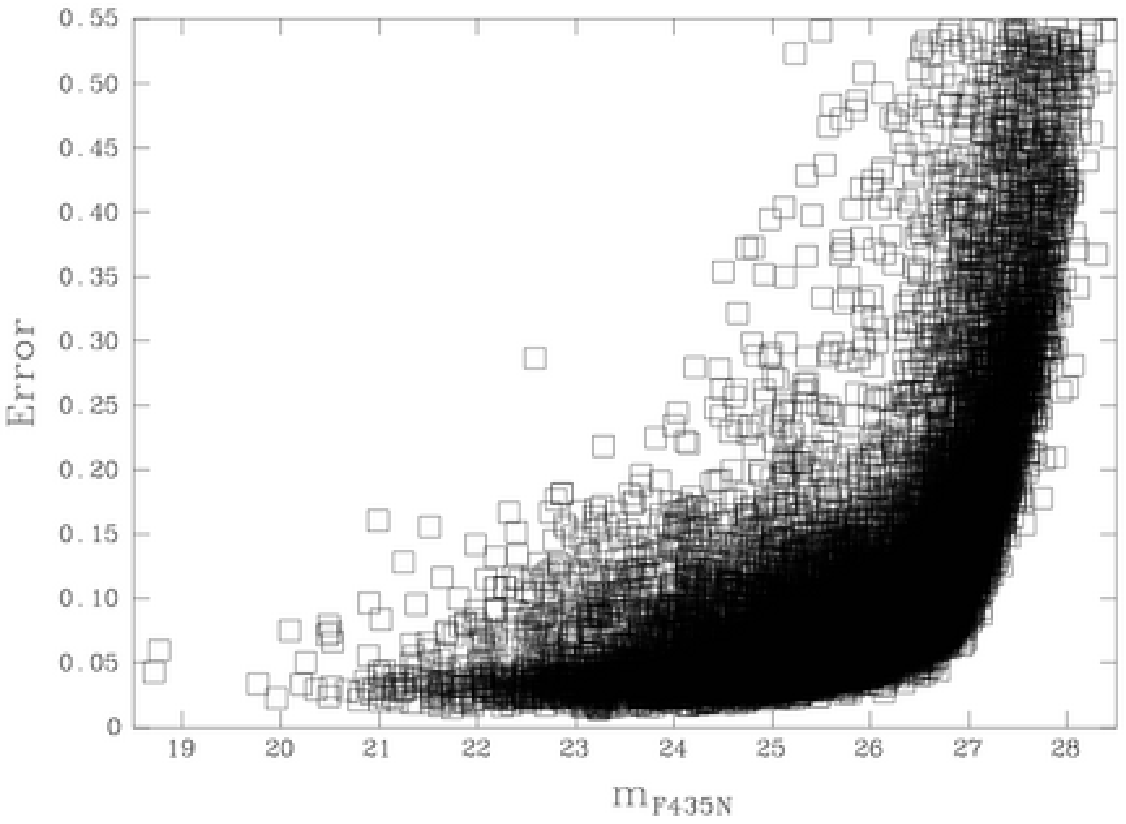} 

\caption{Photometric errors as a function of zero-point corrected, apparent (Vega) magnitude for all sources found in the ACS/WFC F435W image of M101-I .}

\label{f435_errors}

\end{figure}

\subsection{Detection Limits}

\label{detection_limits}

In order to assess the completeness of our M101 survey we must determine to what depth our images probe. Following the method of \citet{bib10} we fit a polynomial to the distribution, where the 100\% detection limit is defined by the point at which the power law deviates from the observed data. Figure \ref{f469_mag_dist} shows the distribution of sources detected in our WFC3/F469N data, indicating that our 100\% detection limit is m$_{F469N}$\,=\,24.3 mag. If we adopt the extinction from \citet{Lee2009} of A(H$\alpha$)\,=\,1.06\,mag, corresponding to A(F469N)\,=\,1.53\,mag following the extinction law from \citet{Cardelli1989}, and adopt the Cepheid distance of 6.4\,Mpc \citep{sha11}, our 100\% completeness detection limit corresponds to M$_{F469N}$\,=\,--6.26\,mag.

% corresponding to an absolute magnitude of M$_{F469N}$\,=\,--4.7 mag, at distance of 6.4\,Mpc and adopting an extinction 

 %only foreground extinction of E(B-V)\,=\,0.007\,mag \citep{sch11}. 

The magnitude distribution of sources in the ACS/F435W data is shown in Figure \ref{f435_mag_dist}, which shows that we sample $\sim$2 magnitudes fainter in the continuum, with a 100\% detection limit of m$_{F435W}$\,=\,26.6\,mag. Again, we adopt the extinction law of \citet{Cardelli1989} so A(F435W)\,=\,1.74\,mag determines a 100\% completeness detection limit of M$_{F435W}$\,=\,--4.17\,mag.

%or M$_{F435W}$\,=\,--2.45\,mag (using the Vega magnitude system).

\begin{figure} 

\centering 

\includegraphics[width=0.7\columnwidth, angle=-90]{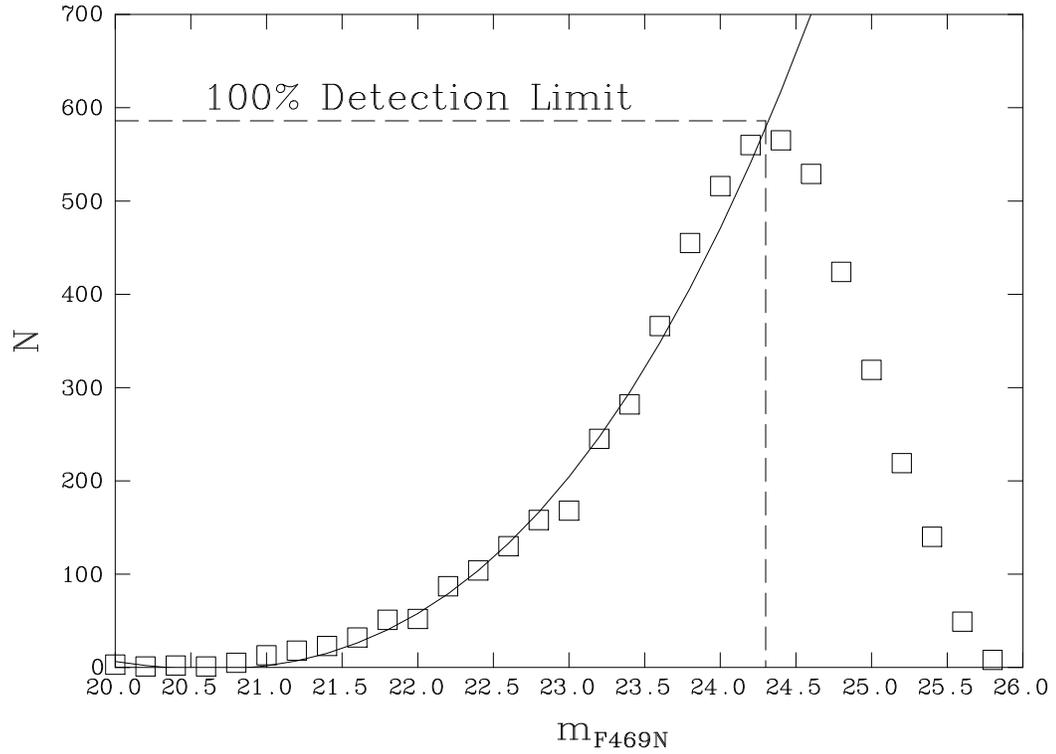} 

\caption{The magnitude distribution of photometric sources identified in the WFC3/F469N image of M101-I using 0.2\,magnitude bins. A 100\% detection limit of m$_{F469N}$\,=\,24.3\,mag is derived from this plot using the solid line which represents a 3rd degree polynomial fit to the brightest sources.}

\label{f469_mag_dist}

\end{figure}

\begin{figure} 

\centering 

\includegraphics[width=0.7\columnwidth, angle=-90]{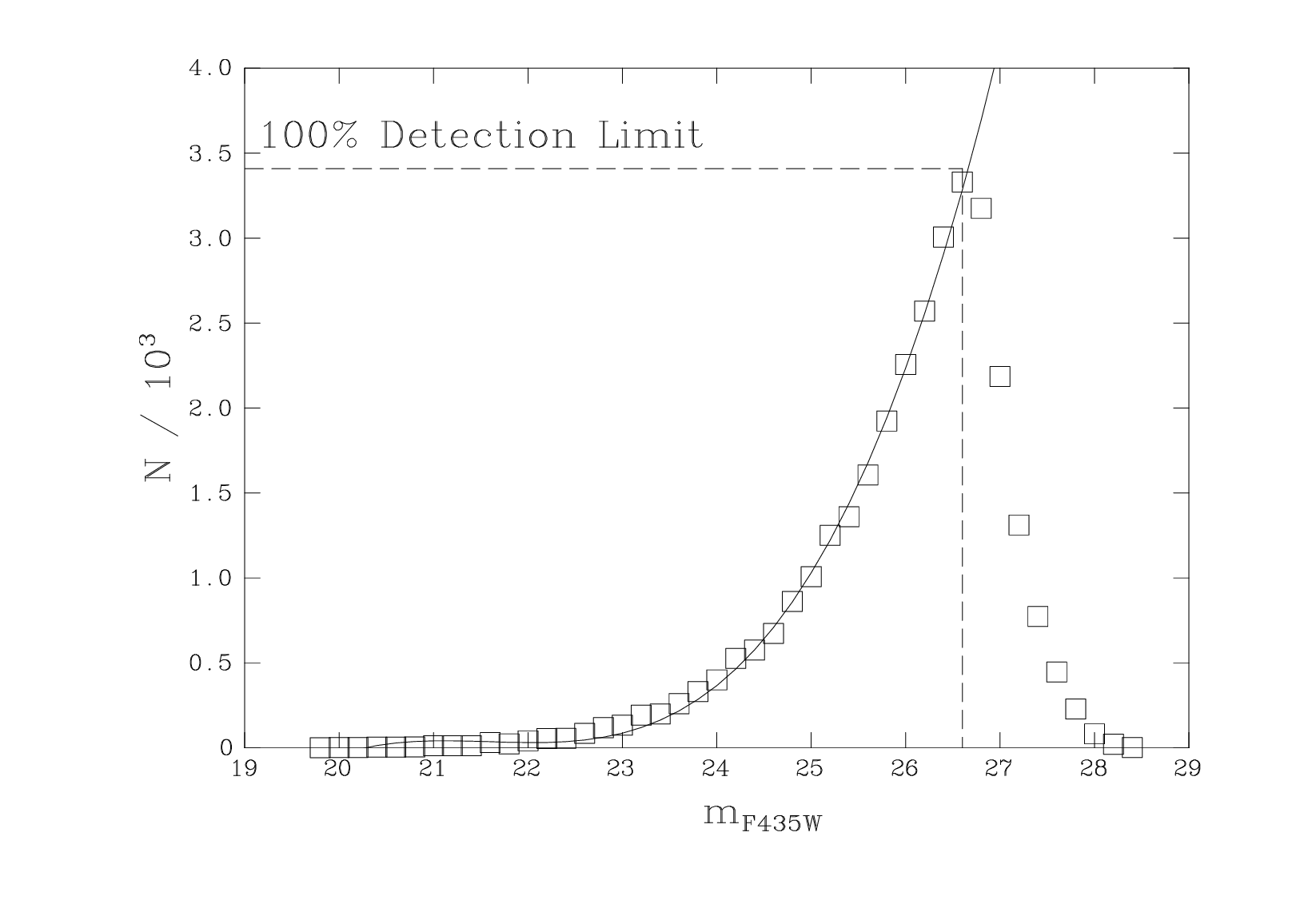} 

\caption{The magnitude distribution of photometric sources identified in the ACS/WFC F435W image of M101-I using 0.2\,magnitude bins. A 100\% detection limit of m$_{F435W}$\,=\,26.6\,mag is derived from this plot using the solid line which represents a 3rd degree polynomial fit to the brightest sources.}

\label{f435_mag_dist}

\end{figure}

\section{Source Selection}

The F469N filter, centered at $\lambda$4693\AA~includes the He\,{\sc ii}$\lambda$4686 emission line, and partially includes the red wing of the C\,{\sc iii}$\lambda$4650 emission lines from carbon--rich WR stars, however the N\,{\sc iii}$\lambda$4640 line from nitrogen--rich, WN, stars will not be detected. In order to identify WR candidates we had to identify sources which were brighter in the narrow-band image relative to the continuum image, in this case the F435W image centered at $\lambda$4297\AA. We note that we also tried creating a continuum image using a combination of the F435W and F555W images, however this did not improve our subtraction and hence we used the F435W and F555W images individually to check the accuracy of our candidates.

Output files from the photometric analysis were merged to match sources in terms of x and y coordinates. For each source m$_{F469N}$--m$_{F435W}$ was determined, where m$_{F469N}$--m$_{F435W}$$\leq$0 indicates an excess at $\lambda$4693\AA. Only sources where the excess was $\geq$3$\sigma$ were considered to be WR candidates. In addition, sources that were only identified in the F469N image, and not in the F435W or F555W images, were also flagged as WR candidates, since it is likely that these stars are faint WR stars with little or no detectable continuum. Figure \ref{m101-i-excess} shows the photometric properties of each WR candidate, with the 100\% detection limit of m$_{F435W}$\,=\,26.6\,mag adopted as the continum magnitude for those candidates only detected in the F469N image.

An efficient way of identifying bonafide WR candidates is via the ``blinking'' method \citep{mof83}, which compares the F469N, F435W and continuum subtracted images in sequence. However, the broad--band F435W filter bandpass is almost 30$\times$ the width of the narrow-band F469N filter, hence the F435W image was scaled to create a narrow-band continuum. Since most stars should have m$_{F469N}$--m$_{F435W}$$=$0 we determined the scale factor which allowed most stars to disappear from the continuum subtracted image. We note that this could affect the detection of WR stars with a low He\,{\sc ii} excess.

The ``blinking'' technique was applied to all of the 3$\sigma$ photometric WR candidates to confirm the He\,{\sc ii}$\lambda$4686 excess and to remove any spurious sources such as cosmic rays, features at the edge of the CCD, and poor photometry in dense, unresolved regions.

In total, after the photometry and blinking processes, 91 WR candidates were identified in just one of our F469N fields (M101-I). Of these, 54 candidates ($\sim$60\%) are detected only in the F469N image. Examples of typical candidates are shown in Figure \ref{blinks} which shows the WR candidate in the F435W, F469N and F469N-F435W$_{scaled}$ filters.

\begin{figure} 

\centering 

\includegraphics[width=0.7\columnwidth, angle=-90]{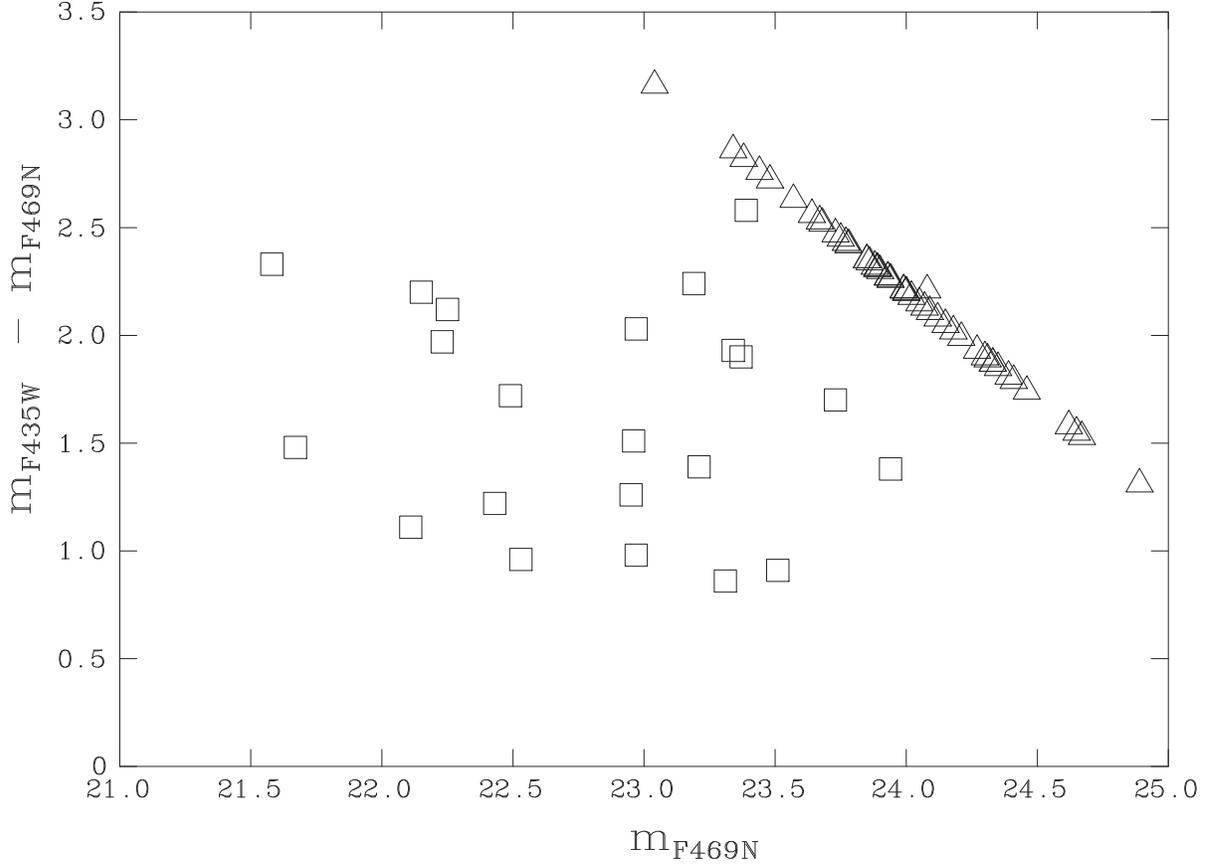} 

\caption{Excess, m$_{F435W}$ -- m$_{F469N}$, versus F469N magnitude for WR candidates in M101-I. Open squares indicate WR candidates with a 3$\sigma$ detection in both F469N and F435W images, which also have an excess of $>$3$\sigma$. Open triangles indicate sources for which there was no detection in the F435W image, so excesses represent lower limits assuming m$_{F435W}$\,=\,26.6\,mag -- our 100\% detection limit. }

\label{m101-i-excess}

\end{figure}

\begin{figure}

\centering

\subfigure[]{\includegraphics[width=0.301\columnwidth]{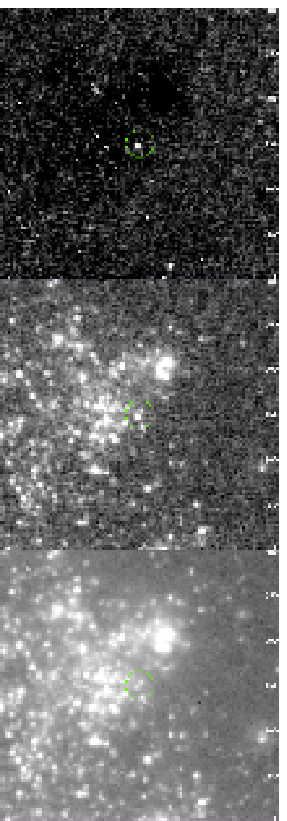}} \subfigure[]{\includegraphics[width=0.3\columnwidth]{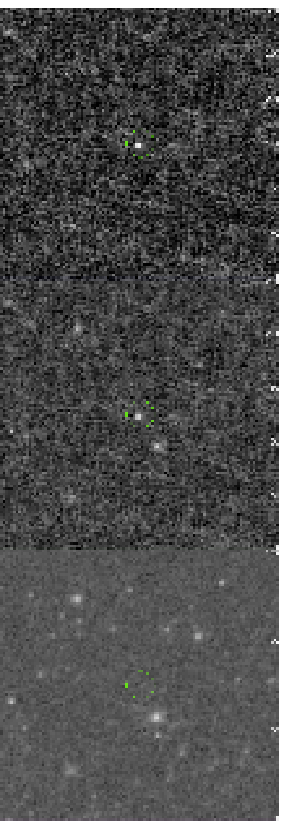}}

\caption{Postage stamp images produced during blinking showing the F435W continuum (bottom), F469N narrow-band (middle) and continuum subtracted image (top) for each WR candidate. Here we show an example of a WR candidate a) detected in the continuum (source \#1026), and b) not detected in the continuum image (source \#639).}

\label{blinks}

\end{figure}

\section{Contamination by Variable Stars}

Since the narrow-band images were obtained in 2010, and the continuum images in 2002, it is possible that stellar variability over the 8 year baseline leads to contamination of our WR candidates with Cepheid and other variable stars.

\citet{sha11} identified Cepheids in 2 fields to determine the distance to M101. They showed that Cepheids have a (V--I) color $\geq$0.5\,mag (their Figure 12) which we can apply to our WR candidates. Of the 37 WR candidates detected in both the F469N and F435W images, and identified to have a $\geq$3$\sigma$ excess in m$_{F469N}$--m$_{F435W}$, 15 are provisionally eliminated ($\sim$40\%) by their red (V--I) $>$0.5\,mag colors. As already notes in Section \ref{detection_limits}, reddening of 0.5\,mag of individual M101 stars, including WR stars, is expected. Hence some of these 15 stars with (V--I)$\geq$0.5\,mag colors may be reddened WR stars. However, to be conservative in our estimates we eliminate them for now. Two of the 22 surviving candidates were not detected in the F814W image, and two other candidates have a V--I $<$-1\,mag;  this is unexpected. However, inspection of the image reveals that these two sources lie in very crowded regions and hence their photometry is likely to be more unreliable than for the other candidates. The (V--I) colors for the 18 candidates which have a $\geq$3$\sigma$ excess and -1$<$(V--I)$<$0.5\,mag are shown in Figure \ref{v-i}, though we emphasize that 22 candidates are not eliminated by the above color test.

Unfortunately, for the WR candidates detected only in F469N, but not in F555W, we cannot calculate a V-I color. However, if we ``blink'' the F469N image with the F814W image we can identify the sources that are also detected in the F814W images. These stars are too red to be WR stars. Of our initial 54 candidates that were not detected in either F435W or F555W, only one is detected in the F814W image. It is is eliminated as a likely red star, reducing the number of WR candidates detected only in F469N, but not in F555W to 53. In the M101-I region our final survey tally is 22 + 53 = 75 candidate WR stars. Their photometry is presented in Table \ref{photometry}.

Blue variable stars could also be a source of contamination in our WR candidate list. We rule out B[e] stars since they do not exhibit any emission lines that lie within the F469N filter bandpass. It is possible that we have identified a few Luminous Blue Variables (LBVs) during outburst, since the continuum of an LBV can vary by up to $\sim$2\,mag. We note that LBVs are rare, with only four confirmed in M33 \citep{cla12}. We expect few LBV to contaminate our WR candidate list, but an independent check, which follows, is prudent.

The HST archive has a rich assortment of M101 images, hence we used ACS/F555W images from 2006 (Proposal ID: 10918, PI Freedman), mid-way between the epochs of the ACS F435W (2002) and WFC3 F469N (2010) imaging, to investigate the impact of variable stars on our survey. The 2006 ACS/F555W imaging covers only a part of the M101-I frame covered by the 2002 narrowband data; 12 of our WR candidates lie within this region. The 2006 data were reduced and analyzed using the same method as described in Section \ref{reduction}, again blinking in all available filters. 12 WR candidates were independently identified using the 2006 imaging, 11 of which are consistent with the analysis using the F555W imaging from 2002. The two remaining candidates, one from each epoch, are candidates that have been identified from the F469N emission and do not have a counterpart in the broad-band imaging. Both candidates are bonafide WR candidates and we conclude that they have most likely been missed through human error in the blinking process. The consistency and lack of variability of the WR candidates found in the two epochs again supports the view that variability is not a significant issue for our survey.

Finally we note that, after the analyses reported here were completed, we obtained Gemini-North spectra (which will be described in the next paper of this series) of over a hundred WR candidates in multiple fields; a high farction do, indeed, turn out to be bona fide WR stars.

\begin{figure}

\centering

\includegraphics[width=0.7\columnwidth, angle=-90]{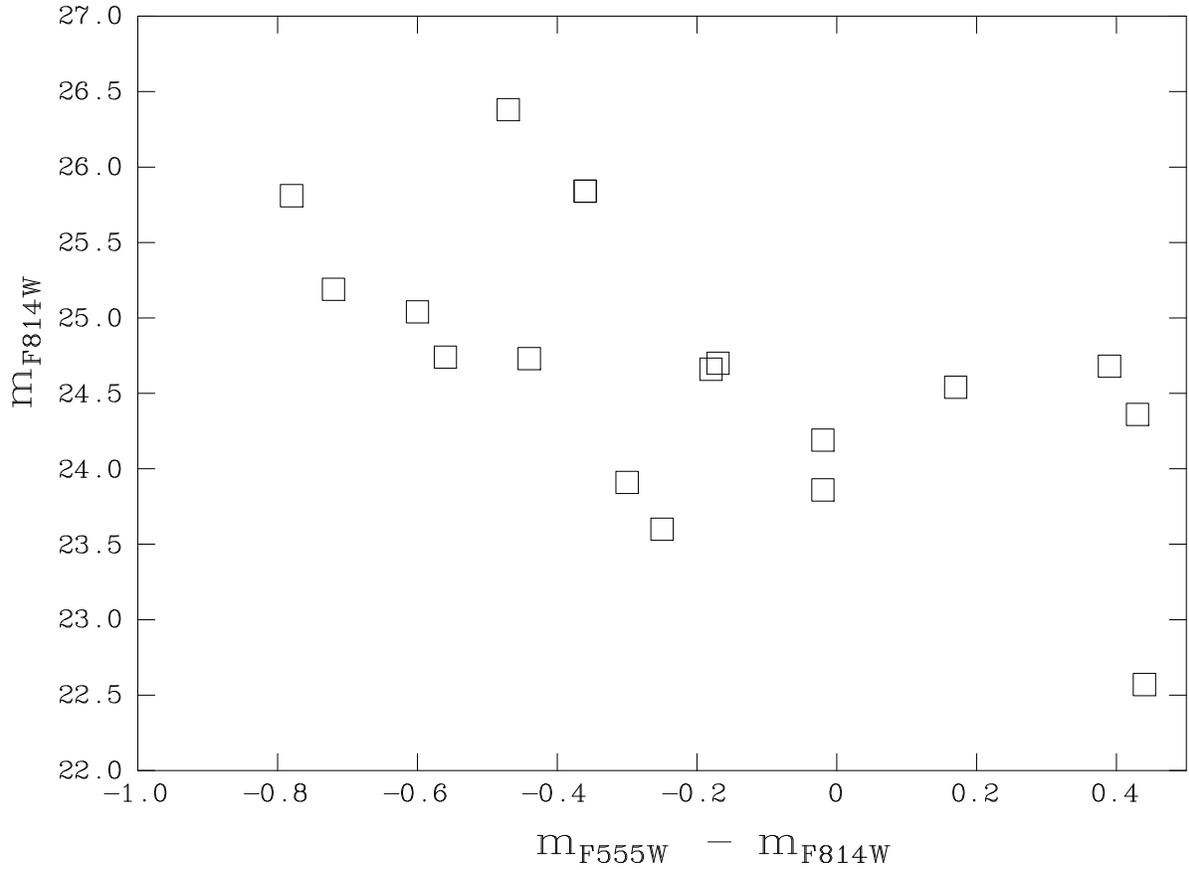}

\caption[]{Here we present the  I (F814W) versus (V-I) of the WR candidates in M101 which have a detection limit in both the F469N and F435W filter images of $>$3$\sigma$. Following \citet{sha11} we use a color cut of  V-I$>$0.5\,mag to rule out contamination by Cepheids. This resulted in the rejection of 15 WR candidates.}

\label{v-i}

\end{figure}

\section{Identifying Red Supergiants}

The archival ACS images also allow us to identify RSG candidates from their (B--V) and (V--I) colors. Based on the colors of RSGs in the LMC we apply color cuts of (B--V)$\geq$1.2\,mag and (V--I)$\geq$1.8\,mag to our field of M101 (B. Davies, priv. communication).  To ensure that we are not contaminated by foreground red giant stars we also insist that the luminosity of each RSG candidate is $\geq$10$^{4.5}$L$_{\odot}$, equivalent to an apparent magnitude of m$_{F814W}\leq$22\,mag. In total we identify 164 RSG candidates in our single M101 field; their photometry is presented in Table \ref{rsg_phot}.

In their study of RSGs in M31, \citet{mas09} suggest a more stringent B--V cut of $\geq$1.6\,mag to remove contamination from foreground stars. However, as they note, adopting this cut removes $\sim$25\% of RSGs already spectroscopically confirmed by \citet{mas98} from their candidate list (their Figure 3). If B--V  $\geq$1.6\,mag is applied to the LMC RSG sample then $\sim$30\% of candidates would be missed, which is consistent with the 35\% reduction in RSG candidates we find if we apply the B--V$\geq$1.6\,mag cut to our M101 sample. The V--I versus B--V colors for our RSG candidates and V--I versus I magnitude for our M101 RSGs are presented in Figure \ref{rsg_a} and \ref{rsg_b}, respectively. These figures depict the candidates which would have been deleted by the more stringent B--V cut as solid squares and the "less stringent" candidates as open squares. Inspection of the open points in Figure \ref{rsg_b} reveals that the photometry of $\sim$7 RSG candidates appears to be inconsistent with the remaining candidates (with m$_{F814W}$$\geq$20.3\,mag or m$_{F555W}$-m$_{F814W}$$\geq$3\,mag). However, we note that the $\delta$m$_{F814W}$\,=\,3\,mag and $\delta$m$_{F555W}$-m$_{F814W}$\,=\,2\,mag range is consistent with that found for the confirmed RSGs in the LMC and as such we do not remove the RSGs from the candidate list.

\begin{figure}

\centering

\subfigure[]{\includegraphics[width=0.5\columnwidth, angle=-90]{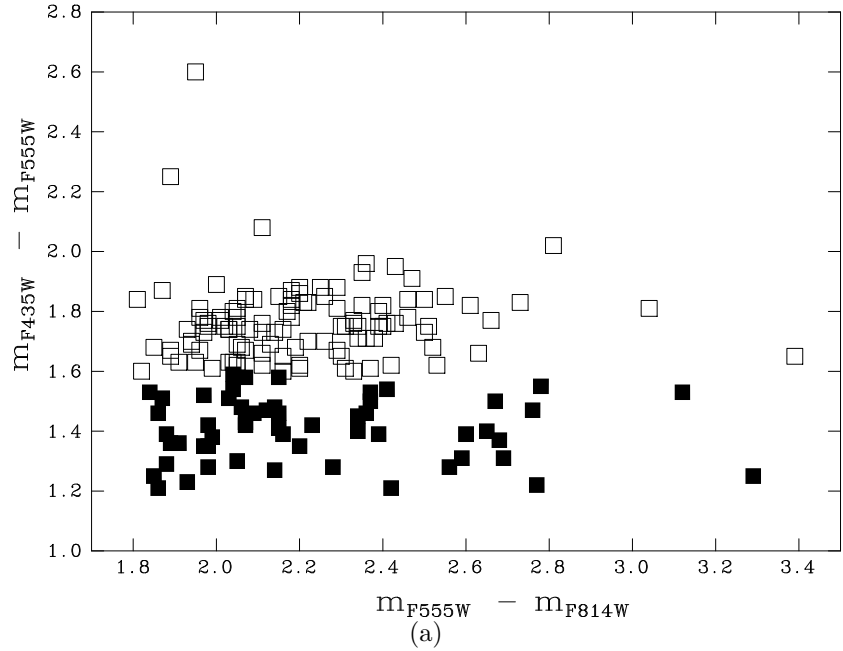} \label{rsg_a}}

\subfigure[]{\includegraphics[width=0.5\columnwidth, angle=-90]{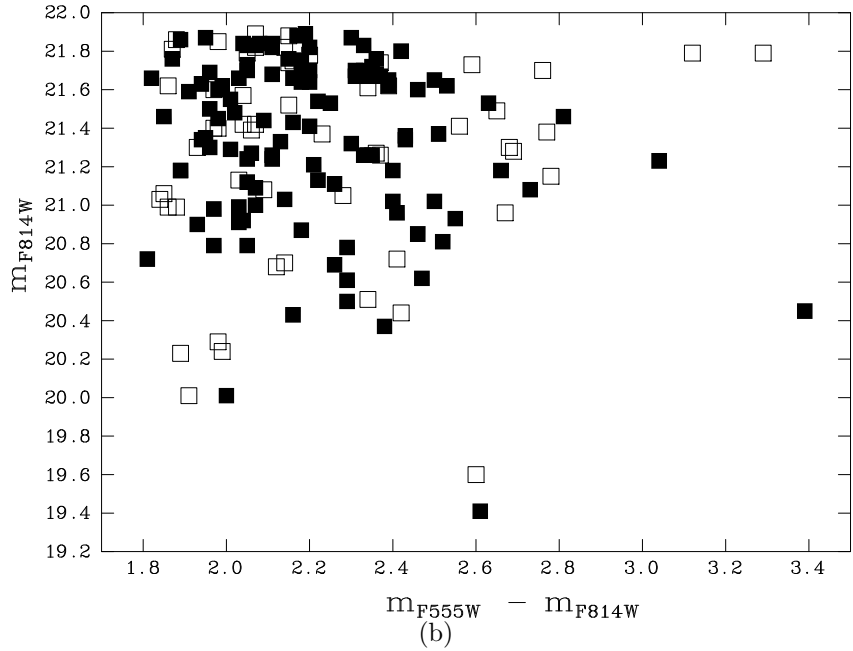} \label{rsg_b}}

\caption{Photometry of Red Supergiant candidates in a single field of M101 showing a) the V--I versus B--V colors and b) V--I versus I-band magnitude for all candidates. The solid squares show sources which are identified as RSGs using a (B--V)$\geq$1.2 cut, while the open squares use the more stringent constraint of (B--V)$\geq$1.6\,mag from \citet{mas09}.}

\end{figure}

\begin{figure}

\centering

\subfigure[]{\includegraphics[width=0.4\columnwidth, angle=-90]{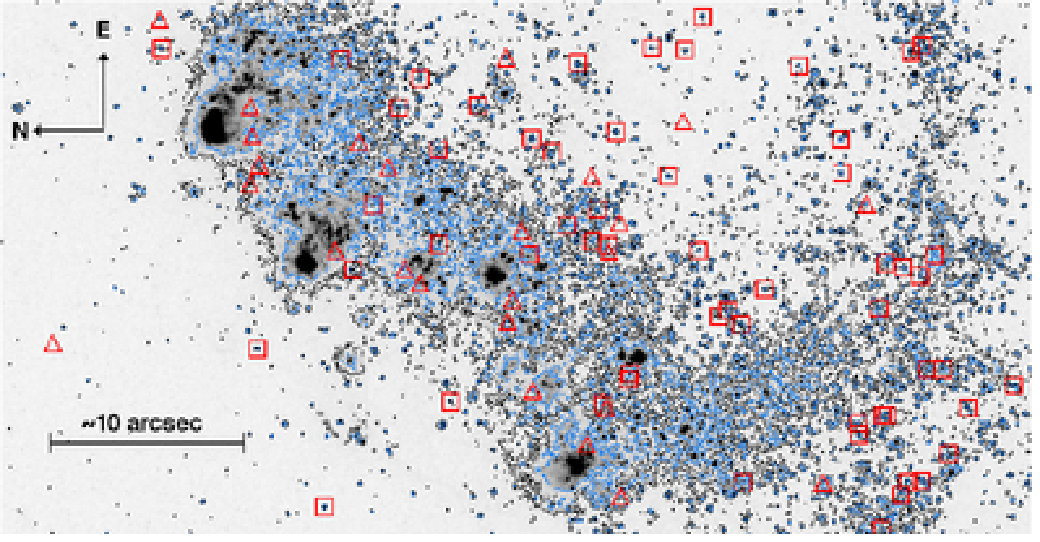}}
\subfigure[]{\includegraphics[width=0.35\columnwidth, angle=35]{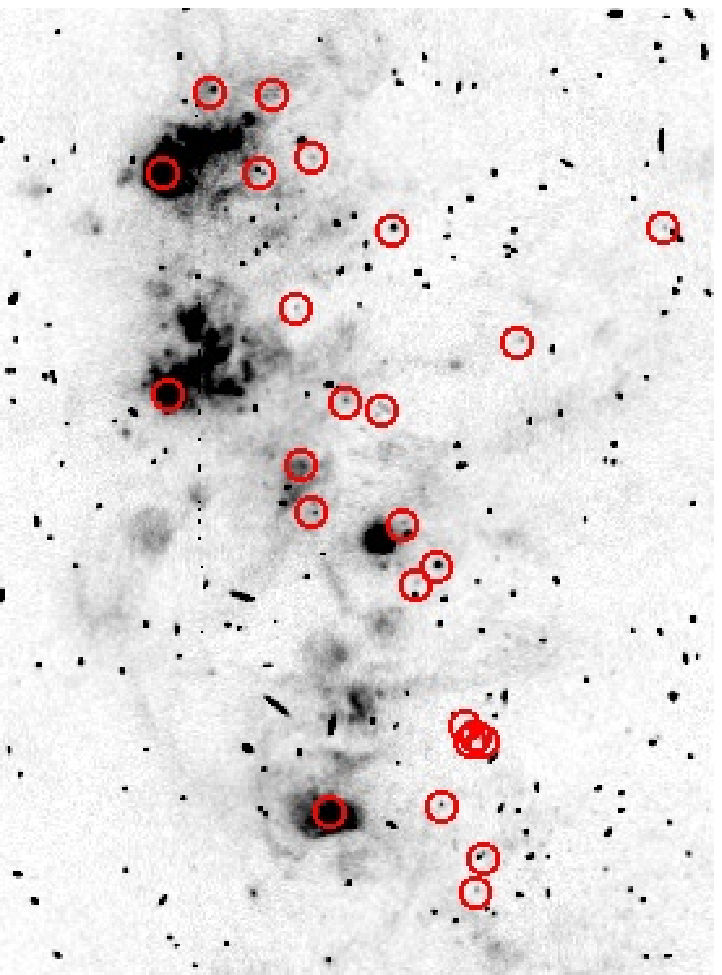}}

\caption{a) Broad band image of the main region of M101-I containing multiple clusters. The contours plotted related to 99\% (blue) and 95\% (black) cuts for the brightness. Wolf-Rayet stars are shown as red triangles and Red Supergiants as red squares - assuming a cut of B-V$\geq$1.6\,mag from \citet{mas09} and b) an archival WFPC2 H$\alpha$ image of the same region with clusters identified from \citet{Che05} plotted as red circles.}

\label{contours}

\end{figure}

In Figure~\ref{contours} we focus on the largest star-forming complex NGC 5462 in the field M101-I, which includes multiple bright H\,{\sc ii} regions from \citet{hod90}, H1159, 1169, 1170 and 1176.  This region has also been studied by \citet{Che05} using HST images and for reference we show an archival HST/WFPC2 H$\alpha$ image of this region, marking the clusters identified by these authors.

RSG and WR candidates are plotted as red squares and red triangles, respectively. The brightest 1\% of all pixels in the ACS field are colored blue, while the next brightest 4\% of all pixels are colored black. 7 WR and 36 RSG candidates are ''isolated'', i.e. not surrounded by black pixels. 6 WR and 30 RSG are largely or entirely surrounded by black pixels. Finally, 12 WR and 5 RSG are surrounded by blue pixels. The corresponding ratios of numbers of WR to RSG candidates are 0.194, 0.2 and 2.4 . The former two are statistically indistinguishable, but the strong clustering of WR stars in the core of the star-forming complex M101-I is suggested by the latter ratio. This suggestion can be made indisputable only with a much larger sample of stars. In later papers in this series we will greatly strengthen these small number statistics with thousands of M101 WR and RSG stars.

\begin{table} \footnotesize \centering \caption[]{Photometry of M101 WR candidates. The RA and DEC of each candidate is taken from the calibration of the F469N/WFC3 image.  Narrow- and broad-band magnitudes are listed for each candidate, unless the object was not detected in that filter. Errors listed are the 1$\sigma$ errors determined by the \textsc{daophot} routine.}

\begin{tabular}{@{\hspace{1mm}}c@{\hspace{3mm}}c@{\hspace{3mm}}c@{\hspace{2mm}}c@{\hspace{3mm}}c@{\hspace{2mm}}c@{\hspace{2mm}}c@{\hspace{2mm}}c@{\hspace{2mm}}c@{\hspace{2mm}}c@{\hspace{2mm}}c@{\hspace{2mm}}c}

  \hline \hline

  RA & DEC & F469N & err & F435W & err & F435W-F469N & err & F555W & err & F814W & err \\

  \hline

14:03:51.802 & +54:21:39.55  & 22.11 & 0.07  &  23.22     & 0.01   & 1.11       & 0.07    & 23.61 & 0.03   & 23.91  & 0.05 \\

14:03:51.378 & +54:21:35.21  & 22.53 & 0.12  &  23.49     & 0.03   & 0.96       & 0.12    & 23.10 & 0.04   &  22.57 & 0.04 \\

14:03:51.722 & +54:21:50.05  & 23.21 & 0.13  &  24.60     & 0.04   & 1.39       & 0.13    & 24.71 & 0.04   & 24.54 & 0.05  \\

14:03:52.347 & +54:21:54.57  & 22.25 & 0.13  &  24.37     & 0.04   & 2.12       & 0.14    & 24.44 & 0.06   &  25.04 & 0.08  \\

14:03:52.755 & +54:21:55.90  & 23.37 & 0.09  &  25.27     & 0.05   & 1.90       & 0.10    & 25.48 & 0.08   & 25.84 & 0.11  \\ 

14:03:54.324 & +54:21:55.99  & 23.34 & 0.12  &  25.27     & 0.05   & 1.93       & 0.10    & 25.48 & 0.08   & 25.84 & 0.11  \\ 

14:03:53.820 & +54:22:03.57  & 23.94 & 0.14  &  25.32     & 0.05   & 1.38       & 0.15    & 25.07 & 0.06   & 24.68 & 0.07   \\ 

14:03:53.690 & +54:22:08.78  & 22.43 & 0.10  &  23.65     & 0.02   & 1.22       & 0.10    & 23.84 & 0.03   & 23.86 & 0.04  \\ 

14:03:53.863 & +54:22:09.11  & 21.58 & 0.12  &  23.91     & 0.04   & 2.33       & 0.13    & 24.18 & 0.06   & 24.74 & 0.07  \\ 

14:03:54.030 & +54:22:09.25  & 23.39 & 0.12  &  25.97     & 0.17   & 2.58       & 0.21    & 25.91 & 0.16   & 26.38 & 0.52  \\ 

14:03:53.344 & +54:21:50.08  & 23.73 & 0.11  &  25.43     & 0.05   & 1.70       & 0.12    & 25.96 & 0.08   & -- & --   \\ 

14:03:53.172 & +54:21:50.63  & 23.19 & 0.12  &  25.43     & 0.05   & 2.24       & 0.12    & 25.96 & 0.08   & -- & --  \\ 

14:03:52.987 & +54:22:00.44  & 21.67 & 0.10  &  23.15     & 0.09   & 1.48       & 0.13    & 23.35 & 0.10   & 23.60 & 0.11  \\ 

14:03:53.072 & +54:22:01.25  & 22.15 & 0.14  &  24.35     & 0.05   & 2.20       & 0.14    & 24.47 & 0.04   & 25.19 & 0.08  \\ 

14:03:53.676 & +54:22:02.08  & 22.23 & 0.14  &  24.20     & 0.03   & 1.97       & 0.14    & 24.10 & 0.03   & 25.28 & 0.07  \\ 

14:03:53.294 & +54:21:55.20  & 22.97 & 0.13  &  23.95     & 0.04   & 0.98       & 0.13    & 24.17 & 0.04   & 24.19 & 0.04  \\ 

14:03:54.942 & +54:22:03.39  & 23.51 & 0.17  &  24.42     & 0.04   & 0.91       & 0.17    & 24.53 & 0.05   & 24.70 & 0.07  \\ 

14:03:53.180 & +54:22:04.81  & 22.49 & 0.17  &  24.21     & 0.06   & 1.72       & 0.18    & 24.29 & 0.08   & 25.49 & 0.20  \\ 

14:03:54.810 & +54:22:08.67  & 22.95 & 0.10  &  24.21     & 0.06   & 1.26       & 0.18    & 24.29 & 0.08   & 24.73 & 0.13  \\ 

14:03:54.551 & +54:22:13.94  & 22.97 & 0.15  &  25.00     & 0.05   & 2.03       & 0.16    & 25.03 & 0.07   & 25.81 & 0.10  \\ 

14:03:36.863 & +54:23:00.19  & 22.96 & 0.14 & 24.47     & 0.04   & 1.51       & 0.15    & 24.79 & 0.05   & 24.36 & 0.04  \\ 

14:03:53.453 & +54:21:37.31  & 23.31 & 0.17   & 24.17     & 0.03   & 0.86       & 0.17    & 24.48 & 0.02   & 24.66 & 0.03  \\ 

14:03:52.091 & +54:21:27.92  & 24.08 & 0.17   & -- & -- & -- & -- & -- & --  & -- & -- \\

14:03:44.772 & +54:21:38.30  & 23.93 & 0.17   & -- & -- & -- & -- & -- & --  & -- & --\\

14:03:47.409 & +54:21:40.92  & 23.99 & 0.18    & -- & -- & -- & -- & -- & --  & -- & --\\

14:03:53.946 & +54:21:46.79  & 24.21 & 0.14   & -- & -- & -- & -- & -- & --  & -- & --\\

14:03:43.399 & +54:22:10.12  & 24.41 & 0.20   & -- & -- & -- & -- & -- & --  & -- & --\\

14:03:46.325 & +54:22:24.73  & 23.88 & 0.11   & -- & -- & -- & -- & -- & --  & -- & --\\

14:03:49.713 & +54:23:06.55  & 23.75 & 0.12   & -- & -- & -- & -- & -- & --  & -- & --\\

14:03:45.544 & +54:23:10.90  & 23.78 & 0.15   & -- & -- & -- & -- & -- & --  & -- & --\\

14:03:48.384 & +54:23:21.04  & 23.94 & 0.10   & -- & -- & -- & -- & -- & --  & -- & --\\

14:03:45.830 & +54:23:23.92  & 24.05 & 0.12    & -- & -- & -- & -- & -- & -- & -- & --\\

%555.92 & 1526.65 & 23.05 & 0.12  &  24.82     & 0.03   & 1.77       & 0.13    & 24.13 & 0.03   & 23.34 & 0.03  \\

%2788.07&  247.68 & 23.25 & 0.12  &  25.08     & 0.04   & 1.83       & 0.13    & 24.29 & 0.03   & 23.40 & 0.03   \\ 

%1225.43 & 851.61 & 23.53 & 0.14  &  25.05     & 0.03   & 1.52       & 0.14    & 24.99 & 0.08   & 24.03 & 0.04  \\ 

%912.19 & 1746.12 & 22.93 & 0.12  &  24.79     & 0.04   & 1.86       & 0.13    & 24.90 & 0.05   & 23.66 & 0.04   \\ 

%3030.25&  753.67 & 23.90 & 0.18  &  25.91     & 0.06   & 2.01       & 0.19    & 24.80 & 0.03   & 23.87 & 0.06  \\ 

%534.22 & 1420.66 & 22.67 & 0.10  &  23.44     & 0.02   & 0.77       & 0.10    & 23.04 & 0.03   & 22.52 & 0.02  \\ 

\end{tabular}

\label{photometry}

\end{table}

\begin{table}

\centering

{Table 1: (continued)} \\

\footnotesize

\begin{tabular}{@{\hspace{1mm}}c@{\hspace{3mm}}c@{\hspace{3mm}}c@{\hspace{2mm}}c@{\hspace{3mm}}c@{\hspace{2mm}}c@{\hspace{2mm}}c@{\hspace{2mm}}c@{\hspace{2mm}}c@{\hspace{2mm}}c@{\hspace{2mm}}c@{\hspace{2mm}}c}

  \hline \hline

RA & DEC & F469N & err & F435W & err & F435W-F469N & err & F555W & err & F814W & err \\

\hline

14:03:43.572 & +54:20:43.93  &   24.15  &     0.17    & -- & -- & -- & -- & -- & -- & -- & --\\

14:03:42.956 & +54:20:56.05  &   23.94  &     0.17    & -- & -- & -- & -- & -- & -- & -- & --\\

14:03:43.450 & +54:21:03.90  &   24.33  &     0.12    & -- & -- & -- & -- & -- & -- & -- & --\\

14:03:42.025 & +54:21:38.77 &   23.38  &     0.13    & -- & -- & -- & -- & -- & -- & -- & --\\

14:03:52.247 & +54:21:50.81 &   23.89  &     0.13    & -- & -- & -- & -- & -- & -- & -- & --\\

14:03:53.618 & +54:21:51.54 &   23.68  &     0.11    & -- & -- & -- & -- & -- & -- & -- & --\\

14:03:47.278 & +54:21:52.04 &   23.67  &     0.18    & -- & -- & -- & -- & -- & -- & -- & --\\

14:03:51.110 & +54:22:05.84 &   24.00  &     0.12    & -- & -- & -- & -- & -- & -- & -- & --\\

14:03:53.577 & +54:22:09.21 &   23.93  &     0.18    & -- & -- & -- & -- & -- & -- & -- & --\\

14:03:43.080 & +54:22:17.27 &   23.99  &     0.11    & -- & -- & -- & -- & -- & -- & -- & --\\

14:03:46.935 & +54:22:18.13 &   24.33  &     0.16    & -- & -- & -- & -- & -- & -- & -- & --\\

14:03:48.954 & +54:22:18.41 &   23.85  &     0.13    & -- & -- & -- & -- & -- & -- & -- & --\\

14:03:52.630 & +54:22:19.51  &   23.73  &     0.15    & -- & -- & -- & -- & -- & -- & -- & --\\

14:03:50.926 & +54:22:31.61 &   23.86  &     0.15   & -- & -- & -- & -- & -- & -- & -- & --\\

14:03:46.177 & +54:22:33.89 &   23.34  &     0.11    & -- & -- & -- & -- & -- & -- & -- & --\\

14:03:36.468 & +54:22:40.98 &   24.46  &     0.19    & -- & -- & -- & -- & -- & -- & -- & --\\

14:03:51.363 & +54:22:42.89  &   24.27  &     0.23   & -- & -- & -- & -- & -- & -- & -- & --\\

14:03:39.703 & +54:23:07.76 &   23.89  &     0.14    & -- & -- & -- & -- & -- & -- & -- & --\\

14:03:43.267 & +54:23:10.18 &   23.64  &     0.11    & -- & -- & -- & -- & -- & --& -- & -- \\

14:03:48.790 & +54:23:30.75 &   23.90  &     0.10    & -- & -- & -- & -- & -- & -- & -- & --\\

14:03:46.573 & +54:20:58.59  &   24.09  &     0.19    & -- & -- & -- & -- & -- & -- & -- & --\\

14:03:40.353 & +54:21:10.67  &   24.31  &     0.30    & -- & -- & -- & -- & -- & -- & -- & --\\

14:03:51.245 & +54:21:17.95  &   23.77  &     0.14    & -- & -- & -- & -- & -- & -- & -- & --\\

14:03:47.905 & +54:21:19.20 &   24.00  &     0.18    & -- & -- & -- & -- & -- & -- & -- & --\\

14:03:45.569 & +54:22:03.20 &   24.02  &     0.16    & -- & -- & -- & -- & -- & -- & -- & --\\

14:03:53.393 & +54:22:22.78  &   23.48  &     0.17    & -- & -- & -- & -- & -- & -- & -- & --\\

14:03:35.495 & +54:22:50.76 &   23.89  &     0.18    & -- & -- & -- & -- & -- & -- & -- & --\\

14:03:43.473 & +54:23:19.11 &   24.89  &     0.31    & -- & -- & -- & -- & -- & -- & -- & --\\

14:03:40.107 & +54:21:14.66  &   23.85  &     0.26    & -- & -- & -- & -- & -- & -- & -- & --\\

14:03:44.015 & +54:21:48.57 &   24.30  &     0.20    & -- & -- & -- & -- & -- & -- & -- & --\\

14:03:40.020 & +54:21:49.87 &   24.62  &     0.48    & -- & -- & -- & -- & -- & -- & -- & --\\

14:03:49.533 & +54:21:57.22 &   24.18  &     0.35    & -- & -- & -- & -- & -- & -- & -- & --\\

14:03:46.586 & +54:22:44.52 &   23.78  &     0.13    & -- & -- & -- & -- & -- & -- & -- & --\\

14:03:43.833 & +54:23:10.16 &   24.00  &     0.15    & -- & -- & -- & -- & -- & -- & -- & --\\

14:03:48.099 & +54:23:39.66 &   24.67  &     0.26    & -- & -- & -- & -- & -- & -- & -- & --\\

14:03:45.029 & +54:21:01.14  &   23.57  &     0.20    & -- & -- & -- & -- & -- & -- & -- & --\\

14:03:52.031 & +54:21:51.80  &   23.04  &     0.12    & -- & -- & -- & -- & -- & -- & -- & --\\

14:03:52.886 & +54:21:55.63  &   23.44  &     0.14    & -- & -- & -- & -- & -- & -- & -- & --\\

14:03:50.001 & +54:23:19.95 &   24.07  &     0.30    & -- & -- & -- & -- & -- & -- & -- & --\\

14:03:40.677 & +54:22:19.23 &   24.65  &     0.20 & -- & -- & -- & -- & -- & -- & -- & --\\

14:03:43.142 & +54:22:34.54 &   24.39  &     0.19 & -- & -- & -- & -- & -- & -- & -- & --\\

14:03:43.161 & +54:22:33.92 &   24.35  &     0.18 & -- & -- & -- & -- & -- & -- & -- & --\\

14:03:38.669 & +54:22:18.74 &   24.12  &     0.16 & -- & -- & -- & -- & -- & -- & -- & --\\

\end{tabular}

\end{table}

%\begin{table}

%\centering

%{Table 1: (continued)} \\

%\footnotesize

%\begin{tabular}{ccclclclclcl}

%\hline \hline

%X & Y & F469N & err & F435W & err & F435W-F469N & err & F555W & err & F814W & err \\

%\hline

%\end{tabular}

%\end{table}

\begin{table}[!]

\footnotesize

\centering

\caption[]{HST/ACS photometry of Red Supergiant candidates in one pointing of M101. In total we identify 164 RSG candidates using color and magnitude cuts provided by B. Davies (priv. communication). Note that all magnitudes presented use the Vega magnitude system, for which zero points were provided by Josh Sokol from the ACS instrument team. Errors listed represent 1$\sigma$ errors which are calculated in \textsc{daophot}.}

\begin{tabular}{ccclclcl}

\\

\hline\hline

RA & DEC & 	F435W &  err  &  F555W & err & F814W & err \\

\hline 

14:03:53.731 & +54:21:53.65 & 25.40 & 0.05 & 24.02 & 0.03 & 21.62 & 0.05 \\

14:03:51.361 & +54:21:13.24 & 25.40 & 0.05 & 23.95  &0.04 & 21.61 & 0.05\\

14:03:52.891 & +54:21:45.49 & 24.94 & 0.04 & 23.48  &0.02 & 21.62 & 0.03\\

14:03:51.615 & +54:21:35.75 & 24.93 & 0.08 & 23.58  &0.03 & 21.60 & 0.03\\

14:03:52.696 & +54:21:38.56 & 25.63 & 0.08 & 24.32 & 0.04 & 21.73 & 0.05\\

14:03:52.978 & +54:21:45.75 & 25.35 & 0.04 & 23.89  &0.03 & 21.74 & 0.04\\

14:03:55.206 & +54:21:54.40 & 25.65 & 0.04 & 24.12 & 0.03 & 21.74 & 0.04\\

14:03:36.509 & +54:22:49.55 & 25.30 & 0.05 & 23.91 & 0.04 & 21.75 & 0.05\\

14:03:51.397 & +54:22:46.20 & 25.33 & 0.05 & 23.98  &0.03 & 21.78 & 0.06\\

14:03:53.121 & +54:21:27.84 & 25.14 & 0.07 & 23.84  &0.03 & 21.79 & 0.04\\

14:03:35.420 & +54:21:11.50 & 26.33 & 0.08 & 25.08  &0.07 & 21.79 & 0.08\\

14:03:53.275 & +54:21:21.52 & 26.43 & 0.07 & 24.91  &0.05 & 21.79 & 0.06\\

14:03:52.172 & +54:21:47.34 & 25.19 & 0.07 & 23.68  &0.04 & 21.81 & 0.06\\

14:03:36.375 & +54:20:48.33 & 25.23 & 0.04 & 23.96 & 0.03 & 21.82 & 0.04\\

14:03:54.280 & +54:21:32.75 & 25.32 & 0.07 & 23.89 & 0.04 & 21.82 & 0.05\\

14:03:53.141 & +54:21:47.40 & 25.25 & 0.07 & 23.84  &0.03 & 21.85 & 0.04\\

14:03:53.387 & +54:21:27.42 & 25.03 & 0.10 & 23.73  &0.05 & 21.86 & 0.06\\

14:03:56.414 & +54:22:32.07 & 25.43 & 0.08 & 24.03  &0.05 & 21.88 & 0.06\\

14:03:51.467 & +54:21:14.10 & 25.38 & 0.05 & 23.96  &0.05 & 21.89 & 0.06\\

14:03:42.430 & +54:23:36.30 & 25.15 & 0.03 & 23.61  &0.02 & 21.57 & 0.04\\

14:03:52.154 & +54:21:39.26 & 25.25 & 0.07 & 23.67 & 0.04 & 21.52 & 0.05\\

14:03:51.011 & +54:21:08.24 & 25.94 &  0.08 &24.46 & 0.04 & 21.70 & 0.05\\

14:03:54.974 & +54:21:53.05 & 25.53 & 0.07 & 24.14  &0.05 & 21.49 & 0.05\\

14:03:53.114 & +54:21:47.13 & 25.06 & 0.04  &23.46  &0.03 & 21.42 & 0.03\\

14:03:52.144 &  +54:21:38.98 & 25.07 & 0.07 & 23.49  &0.03 & 21.42 & 0.04\\

14:03:34.066 & +54:22:47.59 & 25.25 & 0.03  & 23.97 & 0.03 & 21.41 & 0.04\\

14:03:52.364 & +54:21:34.34 & 24.93 & 0.07  &23.45 & 0.02 & 21.39 & 0.03\\

14:03:52.121 & +54:21:41.35 & 24.65 & 0.04  &23.37 & 0.03 & 21.40 & 0.04\\

14:03:41.571 & +54:23:52.62 & 24.89 & 0.04 & 23.37 & 0.03 & 21.40 & 0.05\\

14:03:36.366 & +54:23:07.19 & 25.37 & 0.07 & 24.15 & 0.06 & 21.38 & 0.09\\

14:03:51.947 & +54:21:47.81 & 25.02 & 0.12  &23.60 & 0.05 & 21.37 & 0.06\\

14:03:51.641 & +54:21:37.61 & 24.62 & 0.10  &23.17 & 0.02 & 21.08 & 0.04\\

14:03:54.354 & +54:21:33.96 & 24.45 & 0.06  &23.23 & 0.03 & 21.30 & 0.04\\

14:03:54.335 & +54:21:33.92 & 25.35 & 0.07 & 23.98 & 0.04 & 21.30 & 0.05\\

14:03:45.593 & +54:22:35.95 & 25.09 & 0.11 & 23.63 & 0.11 & 21.27 & 0.16\\

14:03:52.214 & +54:21:28.58 & 25.27 & 0.04 & 23.96 & 0.06 & 21.28 & 0.07\\

14:03:50.700 & +54:21:17.50 & 25.13 & 0.06 & 23.63 & 0.06 & 21.26 & 0.07\\

14:03:51.731 & +54:21:37.81 & 25.49 & 0.04 & 23.94 & 0.02 & 21.15 & 0.02\\

\end{tabular}

\label{rsg_phot}

\end{table}

\begin{table}[!]

\centering

{Table 2: (continued)} \\

\footnotesize

\begin{tabular}{ccclclcl}

\hline\hline

RA  & DEC & 	F435W &  err  &  F555W & err & F814W & err \\

\hline 

14:03:51.696 & +54:21:18.03 	& 	24.66 & 0.05 & 23.15 & 0.06 & 21.13 & 0.09\\

14:03:52.707 & +54:21:48.57 	& 	24.40 & 0.06 & 22.87 & 0.05 & 21.03 & 0.06\\

14:03:51.869 & +54:21:33.31 	& 	24.61 & 0.04 & 23.34 & 0.12 & 21.05 & 0.13\\

14:03:54.266 & +54:22:09.33 	& 	24.16 & 0.06 & 22.91 & 0.04 & 21.06 & 0.06\\

14:03:52.993 & +54:21:54.36 	& 	24.06 & 0.10 & 22.85 & 0.10 & 20.99 & 0.10\\

14:03:36.424 & +54:21:20.63  	& 	24.25&  0.03  &22.87 & 0.03 & 20.99 & 0.04\\

14:03:53.214 & +54:21:58.09 	& 	24.32 & 0.07  &22.84  &0.05 & 20.70 & 0.06\\

14:03:53.119 & +54:21:40.82 	& 	24.67 & 0.03  &23.13  &0.02 & 20.72 & 0.03\\

14:03:51.525 & +54:21:21.18 	& 	24.27 & 0.02  &22.80  &0.02 & 20.68 & 0.03\\

14:03:51.344 & +54:21:00.17 	& 	24.25 & 0.09  &22.85  & 0.06 & 20.51 & 0.07\\

14:03:52.993 & +54:21:54.36 	& 	24.06 & 0.10  &22.85  &0.10 & 20.44 & 0.10\\

14:03:35.731 & +54:22:49.13  	& 	23.59 & 0.03 & 22.20 & 0.03 & 19.60 & 0.04\\

14:03:53.537 & +54:21:59.31 	& 	23.27 & 0.08  &21.91  &0.03 & 20.01 & 0.04\\

14:03:53.403 & +54:21:59.38 	& 	23.48 & 0.05 & 22.12  &0.03 & 20.23 & 0.05\\

14:03:54.224 & +54:21:51.88 	& 	23.61 & 0.06 & 22.23  &0.03 & 20.24 & 0.04\\

14:03:52.916 & +54:21:50.23 	& 	23.63 & 0.03 & 22.28 & 0.03 & 20.29 & 0.05\\

14:03:56.477 & +54:22:11.98 	& 	23.84 & 0.03 & 22.02  &0.04 & 19.41 & 0.05\\

14:03:56.795 & +54:21:54.97 	& 	23.90 & 0.03 & 22.02  &0.03 & 20.01 & 0.04\\

14:03:41.464 & +54:23:35.21  	& 	24.46 & 0.05  & 22.75 & 0.03 & 20.37 & 0.04\\

14:03:52.410 & +54:21:49.64 	& 	24.34 & 0.09 & 22.60  & 0.03 & 20.43 & 0.04\\

14:03:44.033 & +54:23:05.94  	& 	25.49 & 0.05  & 23.84 & 0.03 & 20.45 & 0.04\\

14:03:52.476 & +54:21:34.12 	& 	24.68 & 0.03 & 22.80 & 0.02 & 20.50 & 0.03\\

14:03:52.475 & +54:21:33.35 	& 	24.57 & 0.04 & 22.90 & 0.04 & 20.61 & 0.04\\

14:03:54.025 & +54:22:01.54 	& 	25.00 & 0.06 & 23.09 & 0.04 & 20.62 & 0.05\\

14:03:52.267 & +54:21:50.93 	& 	24.64 & 0.07 & 22.94 & 0.03 & 20.69 & 0.05\\

14:03:53.067 & +54:22:03.94 	& 	24.36 & 0.04  &22.53  &0.03 & 20.72 & 0.04\\

14:03:55.499 & +54:21:35.99 	& 	24.88 & 0.04  &23.07  &0.03 & 20.78 & 0.04\\

14:03:53.845 & +54:21:54.61 	& 	24.53 & 0.05  &22.84  &0.03 & 20.79 & 0.04\\

14:03:54.357 & +54:21:34.98 	& 	24.48 & 0.05  &22.76  &0.04 & 20.79 & 0.06\\

14:03:51.656 & +54:22:05.41 	& 	25.01 & 0.05  &23.33  &0.03 & 20.81 & 0.05\\

14:03:53.157 & +54:21:54.75 	& 	25.15 & 0.10 & 23.31 & 0.09 & 20.85 & 0.13\\

14:03:53.882 & +54:21:50.27 	& 	24.90 & 0.03  &23.06  &0.02 & 20.87 & 0.03\\

14:03:52.740 & +54:21:43.79 	& 	24.57 & 0.03  &22.83  &0.03 & 20.90 & 0.03\\

14:03:52.598 & +54:22:08.88 	& 	24.68 & 0.04  &22.94  &0.03 & 20.91 & 0.03\\

14:03:51.297 & +54:21:19.64 	& 	24.76 & 0.05  &22.96  &0.04 & 20.92 & 0.05\\

14:03:36.903 & +54:22:09.95  	& 	25.33 & 0.04  & 23.49 & 0.04 & 20.93 & 0.05\\

14:03:50.700 & +54:21:17.50  	& 	25.13 & 0.06 & 23.37 & 0.04 & 20.96 & 0.05\\

14:03:51.813 & +54:21:34.43 	& 	24.72 & 0.04 & 22.95 & 0.03 & 20.98 & 0.04\\

14:03:52.838 & +54:21:36.65 	& 	24.76 & 0.05  &23.02  &0.03 & 20.99 & 0.04\\

14:03:53.105 & +54:21:36.27 	& 	24.93 & 0.03  &23.08  &0.02 & 21.00 & 0.02\\

14:03:54.367 & +54:21:46.68         & 	25.23 & 0.04  &23.42  &0.02 & 21.02 & 0.03\\

14:03:50.964 & +54:21:07.60 	& 	25.37 & 0.06 & 23.52  & 0.03 & 21.02 & 0.04\\

\end{tabular}

\end{table}

\begin{table}

\centering

{Table 2: (continued)} \\

\footnotesize

\begin{tabular}{ccclclcl}

\hline\hline

RA  & DEC & 	F435W &  err  &  F555W & err & F814W & err \\

\hline

14:03:56.166 & +54:21:59.42 &  24.90 & 0.04  & 23.17  &0.04 & 21.03 & 0.05\\

14:03:51.522 & +54:21:36.49 &  25.64 & 0.05 & 23.81 & 0.04 & 21.08 & 0.04\\

14:03:54.866 & +54:21:54.63 &  25.00 & 0.05  &23.16  &0.03 & 21.09 & 0.04\\

14:03:52.377 & +54:21:29.64 &  25.22 & 0.03  &23.37  &0.04 & 21.11 & 0.05\\

14:03:53.444 & +54:21:14.05 &  24.97 & 0.04 &  23.17 & 0.03 & 21.12 & 0.05\\

14:03:51.733 & +54:21:35.50 &  25.18 & 0.07  &23.35  &0.09 & 21.13 & 0.12\\

14:03:52.200 & +54:21:36.24 & 25.33 & 0.07    & 23.07 & 0.03&  21.18 & 0.04\\

14:03:52.195 & +54:21:36.42 & 24.74 & 0.03 & 23.07 & 0.03 & 21.18 & 0.04\\

14:03:52.744 & +54:21:20.43 & 25.61 & 0.05  &23.84  &0.04 & 21.18 & 0.05\\

14:03:51.808 & +54:21:35.15 & 25.24 & 0.03  &23.42  &0.02 & 21.21 & 0.03\\

14:03:51.182 & +54:21:57.21 & 26.08 & 0.07  &24.27  &0.03 & 21.23 & 0.05\\

14:03:54.198 & +54:22:00.43 & 24.92 & 0.03  &23.29  &0.02 & 21.24 & 0.03\\

14:03:53.621 & +54:21:47.58 & 25.09 & 0.04 & 23.36 & 0.03 & 21.24 & 0.04\\

14:03:54.963 & +54:21:55.03 & 25.34 & 0.04  &23.60  &0.03 & 21.26 & 0.04\\

14:03:54.378 & +54:22:13.81 & 25.54 & 0.05  &23.61  &0.03 & 21.26 & 0.04\\

14:03:50.700 & +54:21:17.50 &  25.13 & 0.06  &23.37  &0.04 & 21.26 & 0.06\\

14:03:54.288 & +54:21:52.26 & 25.00 & 0.06  &23.32  &0.02 & 21.27 & 0.03\\

14:03:51.808 & +54:21:43.68 & 25.08 & 0.06  &23.31  &0.02 & 21.29 & 0.02\\

14:03:42.360 & +54:23:48.05 & 25.04 & 0.08 & 23.26 & 0.06 & 21.30 & 0.08\\

14:03:44.006 & +54:23:36.89 & 25.37 & 0.10  &23.62  &0.12 & 21.32 & 0.18\\

14:03:55.025 & +54:22:27.55 & 25.15 & 0.06  &23.46  &0.03 & 21.33 & 0.04\\

14:03:51.405 & +54:21:41.58 & 24.98 & 0.03  &23.28  &0.01 & 21.34 & 0.02\\

14:03:52.443 & +54:21:23.97 & 25.73 & 0.06  &23.77  &0.04 & 21.34 & 0.05\\

14:03:54.388 & +54:21:34.39 & 25.89 & 0.07  &23.30  &0.03 & 21.35 & 0.04\\

14:03:52.154 & +54:21:37.76 & 25.54 & 0.05 & 23.78 & 0.03 & 21.36 & 0.04\\

14:03:38.802 & +54:21:47.99 & 25.63 & 0.06  &23.88  &0.03 & 21.37 & 0.04\\

14:03:56.427 & +54:22:04.82 & 25.48 & 0.05  &23.60  &0.03 & 21.41 & 0.05\\

14:03:54.025 & +54:21:42.82 & 25.19 & 0.04  &23.59  &0.03 & 21.43 & 0.04\\

14:03:50.921 & +54:21:22.55 & 25.37 & 0.05  &23.53  &0.03 & 21.44 & 0.04\\

14:03:52.789 & +54:21:44.94 & 25.20 & 0.03  &23.44  &0.01 & 21.45 & 0.02\\

14:03:53.838 & +54:21:38.62 & 26.29 & 0.08 & 24.27 & 0.03 & 21.46 & 0.03\\

14:03:52.447 & +54:21:49.52 & 24.99 & 0.06  &23.31  &0.05 & 21.46 & 0.06\\

14:03:54.564 & +54:21:19.97 & 25.25 & 0.05  &23.50  &0.02 & 21.48 & 0.07\\

14:03:53.208 & +54:21:50.69 & 25.27 & 0.05  &23.46  &0.03 & 21.50 & 0.05\\

14:03:54.040 & +54:21:57.44 & 25.67 & 0.06  &23.78  &0.03 & 21.53 & 0.04\\

14:03:53.151 & +54:21:33.80 & 25.82 & 0.05  &24.16  &0.05 & 21.53 & 0.06\\

14:03:54.772 & +54:22:01.28 & 25.45 & 0.05 & 23.76 & 0.04 & 21.54 & 0.05\\

14:03:53.027 & +54:21:34.49 & 25.34 & 0.05  &23.56  &0.03 & 21.55 & 0.04\\

14:03:53.238 & +54:21:51.47 & 25.13 & 0.07  &23.50  &0.03 & 21.59 & 0.04\\

14:03:51.968 & +54:21:33.02 & 25.84 & 0.07  &24.06  &0.03 & 21.60 & 0.04\\

14:03:52.200 & +54:21:36.24 & 25.33 & 0.07  &23.58  &0.07 & 21.60 & 0.09\\

14:03:35.520 & +54:22:17.92 & 25.22 & 0.07  &23.61  &0.02 & 21.62 & 0.04\\

14:03:53.763 & +54:21:53.60 & 25.77 & 0.07  &24.15  &0.04 & 21.62 & 0.06\\

\end{tabular}

\end{table}

\begin{table}

\centering

{Table 2: (continued)} \\

\footnotesize

\begin{tabular}{ccclclcl}

\hline\hline

X  & Y & 	F435W &  err  &  F555W & err & F814W & err \\

\hline 

14:03:54.566 & +54:21:45.77 	& 	25.26 & 0.04  &23.57  &0.02 & 21.63 & 0.03\\

14:03:56.658 & +54:22:22.04         &	25.63 & 0.05  &23.82  &0.03 & 21.64 & 0.04\\

14:03:52.822 & +54:21:44.51 	& 	 25.45 & 0.04  &23.84  &0.03 & 21.64 & 0.04\\

14:03:54.379 & +54:21:48.41 	& 	 25.84 & 0.06  &24.04  &0.02 & 21.65 & 0.03\\

14:03:49.694 & +54:21:13.17  	& 	 25.89 & 0.07 & 24.15 & 0.04 & 21.65 & 0.05\\

14:03:54.057 & +54:22:02.87 	&      25.09 & 0.05 & 23.48 & 0.03 & 21.66 & 0.04\\

14:03:54.617 & +54:22:56.56 	&      25.32 &0.04 & 23.69 & 0.02  &21.66 & 0.03\\

14:03:37.847 & +54:22:02.26  	&      25.47 & 0.05 & 23.81 & 0.03 & 21.66 & 0.04\\

14:03:43.282 & +54:22:26.49  	&      25.72 & 0.05 & 24.01 & 0.02 & 21.67 & 0.03\\

14:03:51.967 & +54:22:53.10 	&      25.58& 0.07 & 23.98 & 0.05 & 21.67 & 0.06\\   

14:03:52.185 & +54:21:18.89 	&      25.65& 0.07 & 24.04 & 0.05 & 21.67 & 0.06\\   

14:03:53.076 & +54:21:35.38 	& 	25.73& 0.06  &23.86  &0.04  &21.68 & 0.04\\   

14:03:53.446 & +54:22:02.87 	& 	25.87& 0.16  &23.79  &0.04  &21.68 & 0.05\\   

14:03:52.282 & +54:21:58.84 	& 	26.00& 0.08  &24.05  &0.04  &21.68 & 0.05\\

14:03:53.326 & +54:21:52.79 	& 	25.32& 0.08  &23.65  &0.06  &21.69 & 0.08\\

14:03:51.768 & +54:21:23.80         & 	25.75 & 0.07 & 23.90 & 0.04 & 21.70 & 0.05\\

14:03:51.756 & +54:21:23.63         &  	25.76 & 0.10 & 24.01 & 0.06 & 21.70 &0.06\\

14:03:56.264 & +54:21:53.28         & 	25.64 & 0.11  &24.04  &0.04  &21.70 &0.05\\

14:03:50.450 & +54:21:54.07         & 	25.53 & 0.06  &23.75  &0.03  &21.70 &0.05\\

14:03:54.868 & +54:21:59.07         & 	25.89 & 0.09  &24.07  &0.03  &21.72 &0.04\\

14:03:53.418 & +54:21:51.24         & 	25.52 & 0.06  &23.77  &0.03  &21.73 &0.04\\

14:03:55.619 & +54:21:56.82         & 	25.72 & 0.06  &23.94  &0.03  &21.75 &0.04\\

14:03:51.115 & +54:21:27.39         &  	25.83 & 0.06 & 24.12 & 0.03 & 21.76 &0.04\\

14:03:53.779 & +54:21:59.47         & 	25.50 & 0.05  &23.63 & 0.04  &21.76 &0.05\\

14:03:39.781 & +54:21:07.18         &      25.76 & 0.07 & 23.91 & 0.03 &21.76 & 0.05\\

14:03:54.274 & +54:21:40.79         & 	25.83 & 0.05  &24.21 & 0.04  &21.80 &0.05\\

14:03:51.490 & +54:21:21.41         & 	25.64 & 0.05  &24.02 & 0.03  &21.82 &0.04\\

14:03:43.061 & +54:22:28.83         & 	25.60 & 0.06 & 23.94 & 0.04 & 21.82 & 0.05\\

14:03:52.248 & +54:21:32.04         & 	25.50 & 0.05 & 23.87 & 0.02  &21.83  &0.03\\

14:03:37.785 & +54:22:13.83         & 	25.93 & 0.04  &24.15  &0.03  &21.83  &0.04\\

14:03:40.408 & +54:23:25.01         &	25.58 & 0.06  &23.91  &0.03  &21.83  &0.05\\

14:03:52.086 & +54:21:37.68         & 	25.66 & 0.06  &23.92  &0.02  &21.84  &0.03\\

14:03:52.948 & +54:21:42.57         & 	25.51 & 0.06  &23.88  &0.03  &21.84  &0.04\\

14:03:51.309 & +54:21:20.38         & 	25.57 & 0.07  &23.95  &0.06 & 21.84 & 0.07\\

14:03:42.129 & +54:23:49.29         & 	25.39 & 0.08 & 23.75 & 0.05  &21.86  &0.06\\

14:03:53.636 & +54:21:38.60 	& 	25.82 & 0.04  &24.17  &0.02  &21.87  &0.03\\

14:03:53.212 & +54:21:59.41 	& 	25.44 &  0.06 &23.82 & 0.04  &21.87  &0.05\\

14:03:54.313 & +54:22:04.51 	& 	25.85 &  0.15 &24.05 & 0.04  &21.88  &0.05\\

14:03:53.181 & +54:21:45.97 	& 	25.76 &  0.05 &24.08 & 0.04  &21.89  &0.05\\

\end{tabular}                                    

\end{table}

\section{Summary and Conclusions}

We describe the motivation for, and data collected by HST to search for the progenitors of type Ib/c supernovae in the nearby giant spiral galaxy M101. The analysis methodology and early results of a search for WR and RSG stars in one HST WFC3 pointing of M101 are reported. 75 WR and 164 RSG candidates are identified. There is a suggestion of clustering of WR candidates in the central core of the largest star-forming complex in the field. Thousands of WR and RSG candidates, and hundreds of spectrographically confirmed WR stars will be reported in future papers in this series.

\acknowledgments

This research is based on NASA/ESA Hubble Space Telescope observations obtained at the Space Telescope Science Institute, which is operated by the Association of Universities for Research in Astronomy Inc. under NASA contract NAS5-26555. JLB and MMS acknowledge the interest and generous support of Hilary and Ethel Lipsitz. AFJM and LD are grateful to NSERC (Canada) and FQRNT (Quebec) for financial assistance. We thank Or Graur for suggestions on displaying the contours shown in Figure \ref{contours}.

\clearpage

%\begin{figure}

%\figurenum{1}

%\epsscale{1.0}

%\plotone{fig1.eps}

%\caption{}

%\end{figure}

\end{document}